%% file: mnras_template.tex
\documentclass[fleqn,usenatbib]{mnras}

\usepackage{newtxtext,newtxmath}

\usepackage[T1]{fontenc}

\DeclareRobustCommand{\VAN}[3]{#2}
\let\VANthebibliography\thebibliography
\def\thebibliography{\DeclareRobustCommand{\VAN}[3]{##3}\VANthebibliography}

\usepackage{graphicx}	
\usepackage{amsmath}	

\usepackage{lipsum}
\usepackage{empheq}
\usepackage{mathrsfs}
\usepackage{textcomp}
\usepackage{natbib}
\usepackage{hyperref}
\usepackage[caption=false]{subfig}
\usepackage[normalem]{ulem}
\usepackage{xcolor}
\usepackage{bm}
\hypersetup{colorlinks, linkcolor={blue}, citecolor={blue}, urlcolor={blue}}

\newcommand{\kmsmpc}{km\,${\rm s}^{-1}\,{\rm Mpc}^{-1}$\,}

\def\la{\langle}

\def\bea{\begin{eqnarray}}
\def\eea{\end{eqnarray}} 

\title[BayeSN $H_0$]{A BayeSN Distance Ladder: $H_0$ from a consistent modelling of Type Ia supernovae from the optical to the near infrared}
\author[S. Dhawan et al.]{
Suhail Dhawan,$^{1}$\thanks{E-mail: suhail.dhawan@ast.cam.ac.uk}
Stephen Thorp,$^{1,2}$
Kaisey S. Mandel,$^{1,3,4}$
Sam M. Ward,$^{1}$
Gautham Narayan,$^{5,6}$ \newauthor
Saurabh W. Jha,$^{7}$
and Thaisen Chant$^{1,8}$
\\
$^{1}$Institute of Astronomy and Kavli Institute for Cosmology, University of Cambridge, Madingley Road, Cambridge CB3 0HA, UK\\
$^{2}$The Oskar Klein Centre, Department of Physics, Stockholm University, AlbaNova University Centre, SE 106 91 Stockholm, Sweden\\
$^{3}$Statistical Laboratory, DPMMS, University of Cambridge, Wilberforce Road, Cambridge, CB3 0WB, UK\\
$^{4}$The Alan Turing Institute, Euston Road, London, NW1 2DB, UK\\
$^{5}$University of Illinois at Urbana-Champaign, 1003 W. Green St., IL 61801, USA \\
$^{6}$Centre for Astrophysical Surveys, National Centre for Supercomputing Applications, Urbana, IL 61801, USA\\
$^{7}$Department of Physics and Astronomy, Rutgers, the State University of New Jersey, 136 Frelinghuysen Road, Piscataway, NJ 08854, USA\\
$^{8}$Department of Physics, Durham University, South Road, Durham DH1 3LE, UK
}

\date{Accepted XXX. Received YYY; in original form ZZZ}

\pubyear{2022}

\begin{document}
\label{firstpage}
\pagerange{\pageref{firstpage}--\pageref{lastpage}}
\maketitle

\begin{abstract}
The local distance ladder estimate of the Hubble constant ($H_0$) is  important  in  cosmology, given the recent tension with the early universe inference. We estimate $H_0$ from the Type Ia supernova (SN~Ia) distance ladder, inferring SN~Ia distances with the hierarchical Bayesian SED model, BayeSN. This method has a notable advantage of being able to continuously model the optical and near-infrared (NIR) SN~Ia  light curves simultaneously. We use two independent distance indicators,  Cepheids or the tip of the red giant branch (TRGB), to calibrate a Hubble-flow sample of 67 SNe~Ia with optical and NIR data.  We estimate $H_0 = 74.82 \pm 0.97$  (stat)  $\pm\, 0.84$ (sys) \kmsmpc when using the calibration with Cepheid distances to 37 host galaxies of 41 SNe~Ia, and $70.92 \pm 1.14$ (stat) $\pm\,1.49$ (sys) \kmsmpc  when using the calibration with TRGB distances to 15 host galaxies of 18 SNe~Ia.  For both methods, we find a low intrinsic scatter $\sigma_{\rm int} \lesssim 0.1$ mag.  We test various selection criteria and do not find  significant shifts in the estimate of $H_0$. Simultaneous modelling of the optical and NIR yields up to $\sim$15\%  reduction in $H_0$  uncertainty compared to the equivalent optical-only cases. With improvements expected in other rungs of the distance ladder,  leveraging joint optical-NIR SN~Ia data can be critical to reducing the $H_0$ error budget.

\end{abstract}




\section{Introduction}

The Hubble constant ($H_0$) describes the present-day expansion rate and sets the absolute distance scale of the universe. The recent measurement of $H_0$ from the local SN~Ia distance ladder, calibrated to Cepheid variables \citep[e.g.][]{riess2021} is in tension with the inference from the early universe inference using the cosmic microwave background \citep[][]{2020A&A...641A...6P}. Such a tension could potentially be a signature of non-standard physics beyond the standard cosmological model and several studies have reviewed the merits of exotic cosmological models to resolve this discrepancy \citep{2020PhRvD.101d3533K,Shah2021,Schonberg2022}. 
The local measurements are based on a calibration of the absolute luminosity of Type Ia supernovae (SNe~Ia) using independent distances to host galaxies of nearby SNe~Ia. This is termed as the ``cosmic distance ladder".
This claimed tension suggests that the universe at present is expanding about 8-10$\%$ faster than predicted assuming the $\Lambda$CDM model, the concordance cosmological scenario. 
Currently, there are internal differences in the local distance ladder estimate of $H_0$. Calibrating the SN~Ia luminosity using the tip of the red giant branch (TRGB) method \citep[e.g.][]{Freedman2021} does not show a significant tension with the early universe inference.  Understanding differences in the individual rungs of the distance ladder is important to understand whether the tension is a sign of novel cosmological physics or unresolved sources of systematic error. 

In this paper, we focus on the SN~Ia rung of the distance ladder. We  implement a new methodology for  determining the distances to SNe~Ia and, using them to infer $H_0$. Conventionally SN~Ia distances have been derived from optical lightcurves, and subsequent cross-checks have been performed in the near infrared \citep[e.g., see][]{Dhawan2018,Burns2018}. More recently, \citet{galbany2022}, have performed a near-infrared-only analysis with the most updated Cepheid calibrator sample, finding consistent results with the estimates in \citet{riess2021}. They apply a Gaussian process regression method to only single filter NIR ($J$ or $H$ band) photometry to obtain SN~Ia apparent magnitudes from only a single waveband. Here, we test the impact of using a data-driven statistical model for SN~Ia spectral-energy distributions, BayeSN \citep{Thorp2021,Mandel2020}, to improve the inference of the SN~Ia distances.  Crucially, BayeSN models  a continuous SED from the optical through to the NIR and  simultaneously fits the optical and NIR lightcurves. It utilises  all available information for each SN~Ia in a wide wavelength range. The BayeSN model has been previously applied to study dust properties of SNe~Ia in context of their host galaxies \citep{Thorp2021,Thorp2022} and for analysing supernova siblings to estimate $H_0$ from multiple SNe~Ia in the same galaxy \citep{Ward2022}. 
The BayeSN spectral energy distribution (SED) model has been constructed simultaneously from the optical to the NIR wavelengths ($\sim 0.35 - 1.8\,\mu$m). In the NIR wavebands, SNe~Ia have been shown to have a  small intrinsic scatter \citep[e.g. see][for earlier works on the uniform behaviour of SNe~Ia in the NIR]{elias1981,Elias1985,meikle2000,krisciunas2004}.  NIR magnitudes at maximum light exhibit small scatter \emph{without} the typical lightcurve shape and colour corrections that are applied in the optical wavelengths \citep[e.g.][]{Wood-Vasey2008,Mandel2009,Folatelli2010, Mandel2011,Barone-Nugent2012, Kattner2012,Weyant2014,Friedman2015,Avelino2019,Johansson2021}.  Simultaneously fitting the optical and NIR lightcurves also enables a more accurate determination of the host galaxy dust extinction \citep[e.g.][]{Krisciunas2000, Krisciunas2007,Mandel2011,Burns2014,Thorp2022}. The BayeSN model can be used to exploit both the low luminosity dispersion of SNe~Ia in the NIR, and improve upon constraints by leveraging the long wavelength baseline of SN~Ia observations. With optical and NIR data, BayeSN has been demonstrated to have lower root-mean-square (RMS) scatter than conventional lightcurve fitting tools, e.g. SNooPy \citep{Burns2011}, SALT2 \citep{Guy2007}. Given the unique capabilities of BayeSN to simultaneously model the optical and NIR to infer more accurate distances, we investigate the impact of the improved distance inference model on cosmological parameter estimation. In this study, we focus on $H_0$ from the local distance ladder.
\begin{figure}
    \centering
    \includegraphics[width=.48\textwidth]{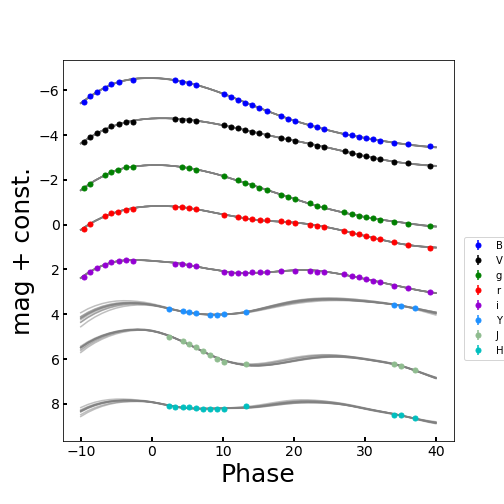}
    \caption{An example BayeSN fit to the multi-wavelength optical-NIR data for SN~2015F \citep{Cartier2014}. BayeSN simultaneously fits all the data from the $B$ to the $H$ band for this object, exploiting the constraining power of the NIR.}
    \label{fig:example_bayesn}
\end{figure}
In addition to current datasets, there are several forthcoming surveys with a large component dedicated to NIR observations, e.g. the Carnegie Supernova Project-II \citep[CSP-II;][]{Phillips2019}, SIRAH \citep{Jha2019} with HST, FLOWS \citep{Muller-Bravo2022}, and the DEHVILS survey using UKIRT \citep{Konchady2022}. Moreover, recent studies of high-$z$ ($0.2 < z < 0.6$) SN~Ia observed in the NIR via the RAISIN program have demonstrated the promise of using the NIR as an independent route to measure dark energy properties \citep{Jones2022:raisin}. Future space missions, e.g. the Roman Space Telescope, with optimised sensitivity in the NIR wavebands are forecast to precisely constrain properties of accelerated expansion \citep{hounsell2018,Rose2021}.
In this paper, we focus on the constraints on $H_0$ from a low-$z$ sample. We describe our dataset and methodology in Section~\ref{sec:data} and present our results in Section~\ref{sec:results}. We discuss the results in context of other studies in the literature and conclude in section~\ref{sec:disc}.

\input{tables/calib_DM_Ceph}
\begin{figure*}
    \centering
    \includegraphics[width=.48\textwidth]{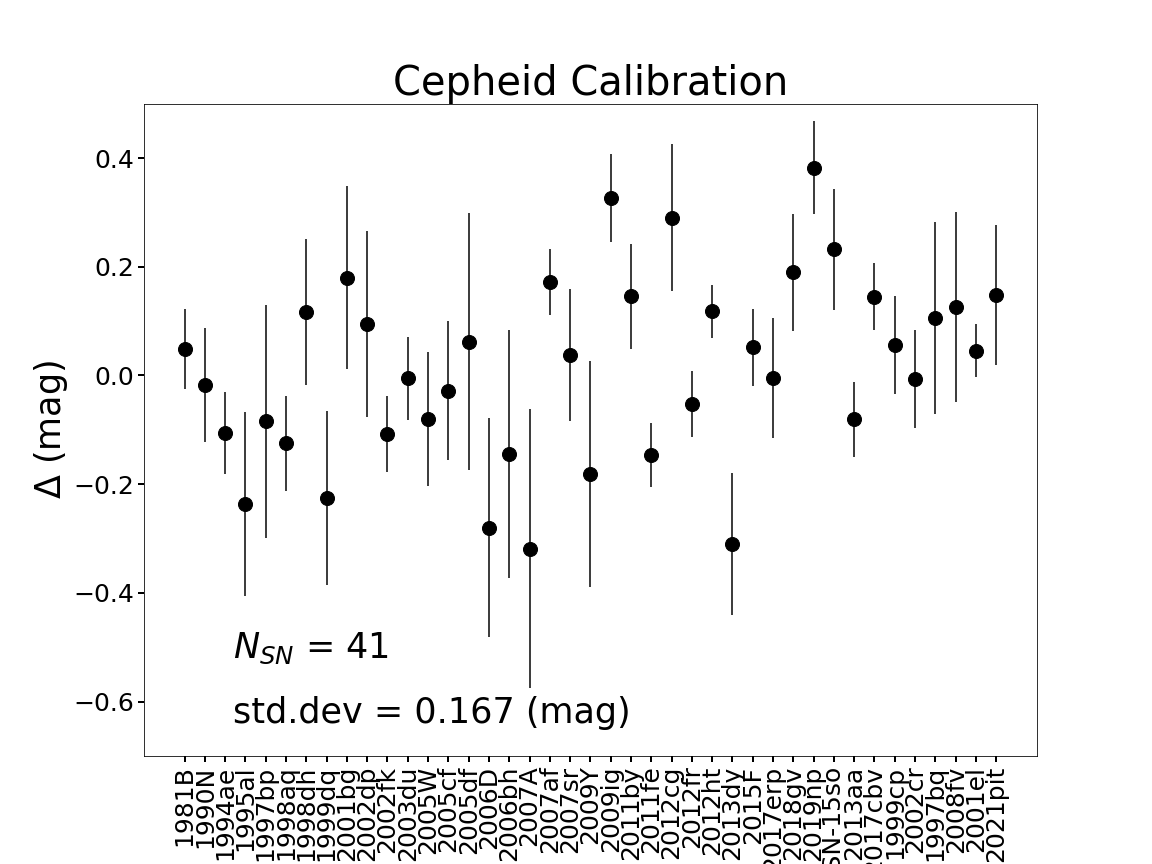}
        \includegraphics[width=.48\textwidth]{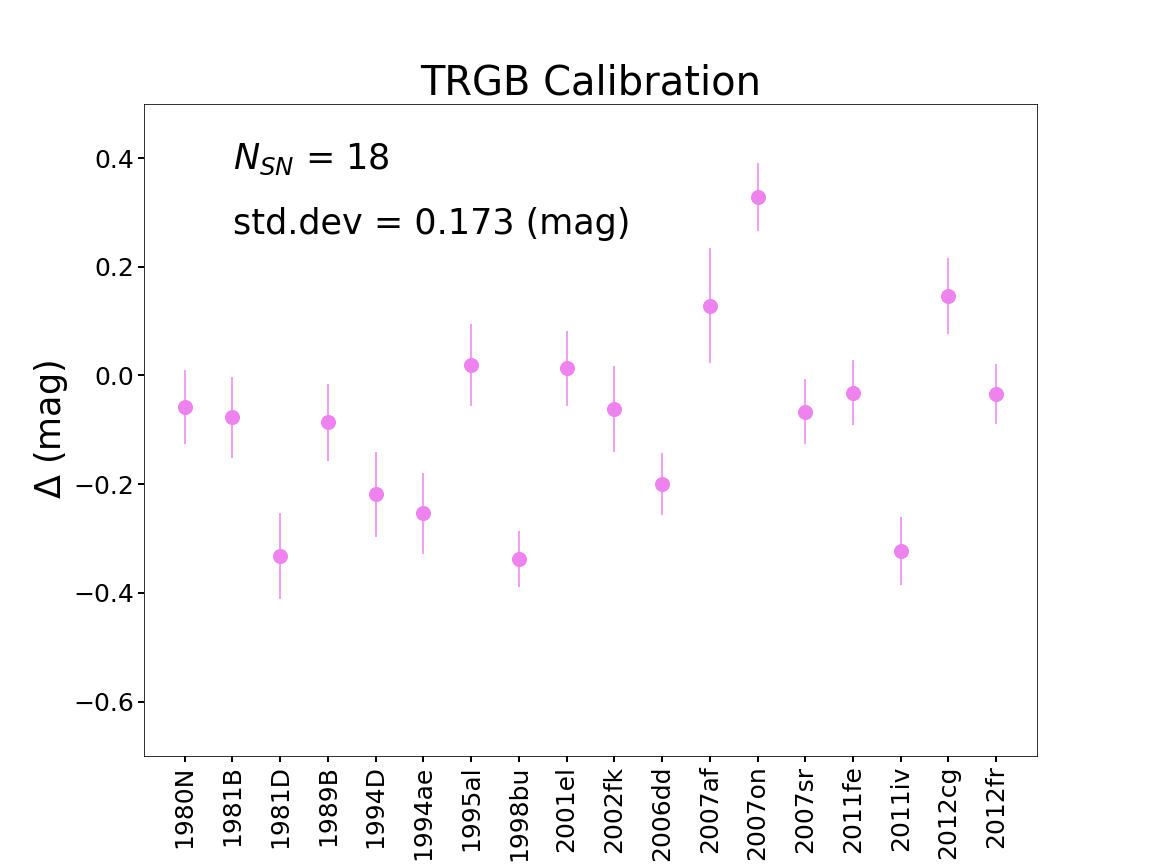}
\caption{Difference, $\Delta = \hat{\mu}_\text{SN} - \hat{\mu}_\text{Cal}$, between the SN~Ia photometric distance and the independent distance to the host galaxy for the calibrator sample, using the Cepheid (left, black) and TRGB calibration of the distance (right, violet) to the SN~Ia host galaxies. The sample standard deviation of these differences is indicated.  We note that the plotted errors do not include the SN intrinsic scatter, however, the intrinsic scatter was added in quadrature in the analysis.  We note that, in each case, the average of the differences corresponds to the differences between the reference scale assumed in BayeSN and the scale set by the independent distance indicator. We also emphasize that this does not impact our final estimate of $H_0$. }
    
    \label{fig:distance_ladder}
\end{figure*}
\section{Data and Methodology}
\label{sec:data}
\begin{figure}
   \centering
    \includegraphics[width=.46\textwidth]{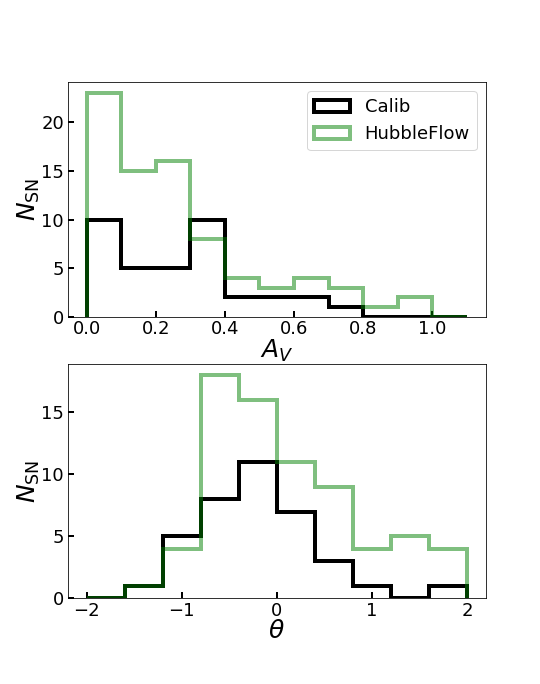}
    \caption{(Left): Comparison of the $A_V$ (top) and $\theta$ (bottom) distribution between the Cepheid calibrator (green) and the Hubble flow samples (red). We note that the calibrator sample has fewer SNe with high $A_V$ and $\theta$ and test the impact of this mismatch on the final $H_0$ (see text for details)}

       \label{fig:av_theta_dist}
\end{figure}
\begin{figure}
   \centering
    \includegraphics[width=.46\textwidth]{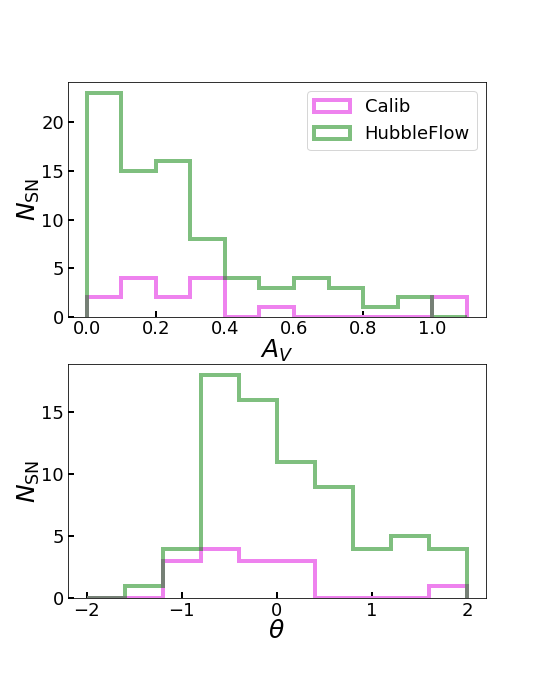}
    \caption{Same as figure~\ref{fig:av_theta_dist} but for the TRGB calibration}
       \label{fig:av_theta_dist_trgb}
\end{figure}

For constraining $H_0$ using SNe~Ia modelled with BayeSN, we need a sample of nearby SNe~Ia with independent distances to their host galaxies (calibrators). The most widely used methods to get distances to nearby galaxies are Cepheid variables \citep{riess2021} and the tip of the red giant branch (TRGB) method \citep{Freedman2021}, both of which have been used for $H_0$ inference with sample of order tens of SN~Ia galaxies. Secondly, we require a sample of SNe~Ia in the Hubble flow ($z > 0.01$). 

For our Cepheid-calibrated sample we use the host galaxy distances from \citet{riess2021}, for a total of 37 host galaxies of 41 SNe~Ia. Out of these 41 SNe~Ia, 15 have NIR data ($\geq 1\, \mu$m). Individual sources for the datasets are detailed in Table~\ref{tab:calib_ceph}. For the 23 new SNe~Ia in the sample of calibrators presented in \citet{riess2021}, we take the data provided\footnote{We fit all the available SNe~Ia with the SED model.  However, we could not obtain an adequate fit to the provided data for SN~2021hpr, so it was omitted. Hence, we use the total of 41 SNe~Ia in 37 host galaxies in the calibrator-sample.} as part of the Pantheon+ data release \citep{brout2021,scolnic2021}. For the TRGB method, we use the sample of 18 galaxies with distances provided \citet{Freedman2019}. A summary of the samples is provided in Table~\ref{tab:calib_ceph} and~\ref{tab:calib_trgb} for the Cepheid and TRGB method, respectively.   
While the current number of SNe~Ia with NIR data is significantly lower than those with optical data, there have been a few dedicated follow-up programs, e.g. the Carnegie Supernova Project \citep[CSP-I;][]{Krisciunas2017}, the Center for Astrophysics (CfA) SN program \citep{Wood-Vasey2008, Friedman2015} as well as programs for follow-up of the Palomar Transient Factory \citep[PTF;][]{Barone-Nugent2012}, intermediate Palomar Transient Factory  \citep[iPTF;][]{Johansson2021} and the SweetSpot survey \citep{Weyant2018}. Since our aim is to model the optical and NIR simultaneously we take the training sample of SNe~Ia from \citet{Mandel2020} with $z \geq 0.01$, a total of 67 SNe, as our fiducial Hubble flow sample. This sample was compiled from the CSP-I and CfA samples and objects from the literature (c.f. Table 1 of \citealt{Mandel2020}). We obtain the heliocentric frame redshifts from \citet{Avelino2019}. 
\begin{figure}
    \centering
    \includegraphics[width=.5\textwidth]{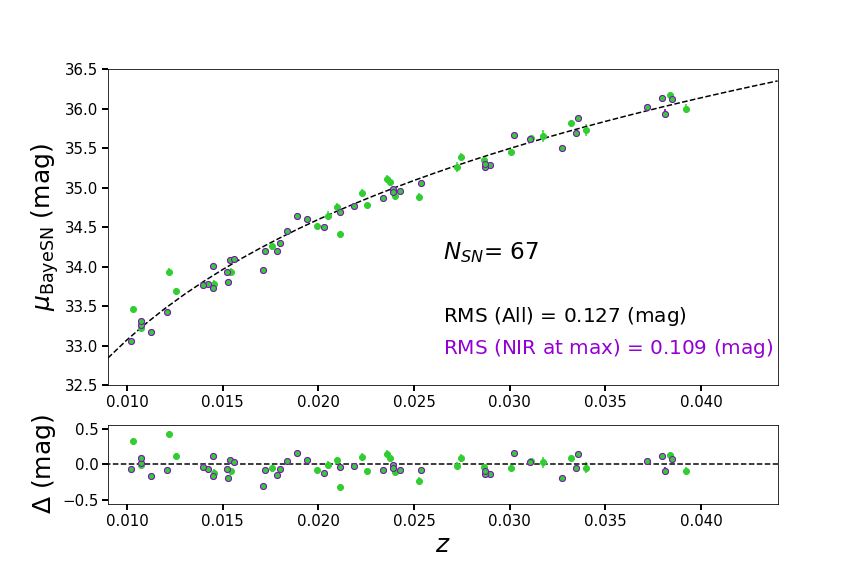}
    \caption{Photometric BayeSN optical+NIR distance estimates versus redshifts for the Hubble flow sample in our analysis. The error bars do not include the $\sigma_{\rm int}$ term.}
    \label{fig:hflow}
\end{figure}

We fit the lightcurves of SNe in the calibrator and the Hubble flow samples using the BayeSN model. We summarise the method here and refer to \citet{Mandel2020,Thorp2021} for details. BayeSN is a hierarchical Bayesian model of continuous time-dependent SN~Ia SEDs, and generalises the previous optical-NIR light curve models of \citet{Mandel2009, Mandel2011}. It models two populations, an intrinsic component and a component for extinction from host galaxy dust. The intrinsic  SED is a  time- and wavelength-dependent function constructed from a functional principal component,  with a coefficient $\theta$ parametrizing the primary light curve shape, and a residual function  with a covariance matrix  derived from model training \citep{Mandel2020,Thorp2021,Ward2022}. The dust component is modelled by the \citet{Fitzpatrick1999} dust law with two parameters ($A_V$, $R_V$), the extinction in the $V$-band and the total-to-selective absorption ratio that parametrizes the steepness of the dust law. 
Following from \citet{Mandel2020}, the fits are all performed in the Stan probabilistic programming language \citep{Carpenter2017,Stan2020}. The joint posterior over all individual and global parameters is sampled using a Hamiltonian Monte Carlo algorithm \citep{Hoffman14,Betancourt16}. 
Previous iterations of the BayeSN models are either trained on $BVriYJH$ \citep{Mandel2020} or $griz$ \citep{Thorp2021} sets of filters. In our analyses both calibrator and Hubble flow samples have SNe~Ia that have measurements in passbands that are in either, and sometimes both sets of filters. Therefore, we implement a new BayeSN model (``W22") that was trained  simultaneously on the optical-NIR $BgVrizYJH$ data of the  combined M20 and T21 training samples, comprising a total of 236 SNe~Ia. Robustness tests for this model and details of the training are presented in a companion paper \citep{Ward2022}. Our lightcurve fits return  estimates of the photometric distance $\mu_{\text{SN}}$, extinction $A_V$ and lightcurve shape $\theta$ parameters.

The model training adopted a reference distance scale with $H_0^{\rm ref} = 73.24 \,{\rm km}\,{\rm s}^{-1} {\rm Mpc}^{-1}$, $\Omega_{\rm M} = 0.28$ and a flat universe \citep{riess2016}. This sets an absolute magnitude normalisation within the model. We emphasize that the estimation of $H_0$ requires a relative comparison of the SNe~Ia in the calibrator and the Hubble flow samples. In our analyses, we use the same absolute  magnitude normalisation for the inferred SN-based photometric distances, for \emph{both} the calibrator and Hubble flow samples. Therefore, any effect of assuming a specific reference distance scale in the training cancels out since it is applied to both the calibrator and Hubble flow samples. Hereafter, we represent the best fit value for the BayeSN photometric distance as $\hat{\mu}_{\rm SN}$ and the associated  fitting uncertainty as $\hat{\sigma}_{\rm SN, fit}$.
\input{tables/calib_DM_TRGB}

For the fitting process we ignore data in filters that are bluer than 0.35 $\mu$m, e.g. the $u$ band or redder than 1.8 $\mu$m, e.g. the $K$-band. Since the W22 model is defined in the phase range $-10$ to $+40$ rest-frame days from $B$-band maximum, we ignore data outside this phase range. For our analyses, we keep the total-to-selective absorption ratio for the extinction correction fixed to the value of $R_V = 2.659$ from the training sample. An example fit to a calibrator object SN~2015F \citep{Cartier2014} is shown in Figure~\ref{fig:example_bayesn}.  For the light curve fit, we set $\sigma_{\rm int}$, the parameter\footnote{This is denoted $\sigma_0$ in \citet{Mandel2020}.} within BayeSN representing the achromatic intrinsic scatter, to 0. Instead, the intrinsic scatter parameters will be inferred simultaneously with the cosmological parameters. This fiducial case is termed as the one scatter model. We also test for consistency between the scatter value derived from the calibrator and Hubble flow samples separately. In the Cepheid calibrator sample, there are 41 SNe~Ia in 37 host galaxies, and in the TRGB calibrator sample, two out of the 15 host galaxies of 18 SNe~Ia host two or more SNe~Ia. For our analysis, we take the weighted mean of the inferred SN~Ia distance to avoid twice counting the host galaxy data.

\begin{figure*}
    \centering
    \includegraphics[width=.48\textwidth]{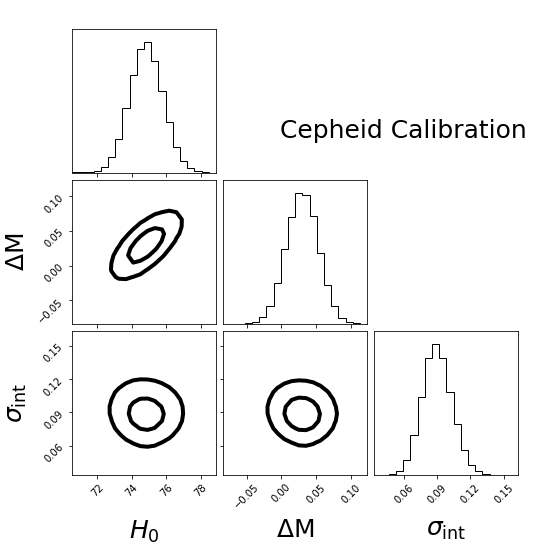}
    \includegraphics[width=.48\textwidth]{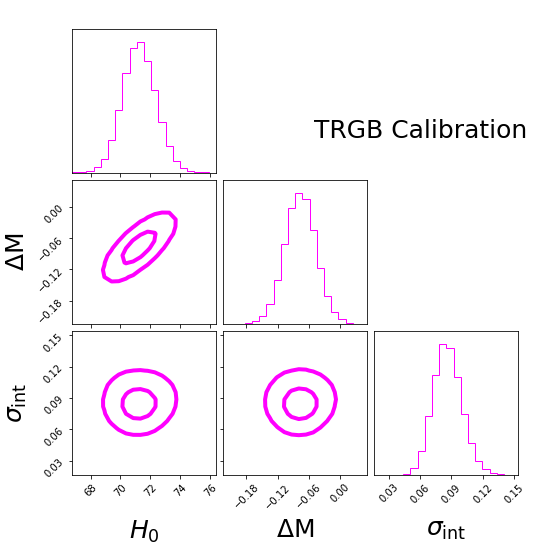}
    \caption{The corner plot showing the posterior distribution for the $H_0$, $M_X$ and $\sigma_{\rm int}$ using the Cepheid (left) and TRGB (right) calibrated distance ladder. With the Cepheid distance ladder we find $H_0 = 74.82 \pm 0.97$ \kmsmpc and for the TRGB we find $H_0 = 70.92 \pm 1.14$ \kmsmpc. We find a small $\sigma_{\rm int} = 0.094 \pm 0.013$ for the Cepheid case and $\sigma_{\rm int} = 0.085 \pm 0.014$ mag for the TRGB case.}
    \label{fig:contour_H0}
\end{figure*}

\section{Analysis and Results} 
\label{sec:results}
In this section, we present the results from fitting the SNe~Ia in the calibrator and Hubble flow samples and inferring $H_0$. A summary of the calibrator SN~Ia fits and the absolute distances from the Cepheid variables is presented in Table~\ref{tab:calib_ceph}. We emphasise again that since the distance ladder is constructed  from the relative measurements of the calibrator and Hubble flow SNe~Ia, the reference distance scale used in the model training is an arbitrary factor that does not impact the final inference on $H_0$,  since it is absorbed into the determination of the absolute magnitude offset ($\Delta M$ as defined below).
\input{tables/hubbleflow}

For the Hubble flow sample we present the redshifts and the fitted SN~Ia distances in Table~\ref{tab:hubbleflow}. 
Typically, the heliocentric corrections to the CMB frame are done using an additive approximation such that
\begin{equation}
    z_{\rm Hel} = z^{+}_{\rm CMB} + z_{\rm Sun}
\end{equation}
However, as noted by \citet{Davis2019, Carr2021}, the correct way to transform heliocentric redshifts is to multiplicatively combine the $(1+z)$ terms, i.e.
\begin{equation}
(1+z_{\rm Hel}) = (1 + z^{X}_{\rm CMB})(1 + z_{\rm Sun})
\end{equation}
which gives the CMB frame redshift as 
\begin{equation}
z^{X}_{\rm CMB} = \frac{(1+z_{\rm Hel})}{(1 + z_{\rm Sun})} - 1
\label{eq:zcmb_trans}
\end{equation}
the difference between the additive and multiplicative formulae is exactly $z_{\rm CMB}z_{\rm Sun}$. While this effect is small at low-$z$, it becomes significant at higher redshifts. For consistency, we transform to CMB frame using equation~\ref{eq:zcmb_trans}.

\begin{figure*}
    \centering
    \includegraphics[width=.48\textwidth]{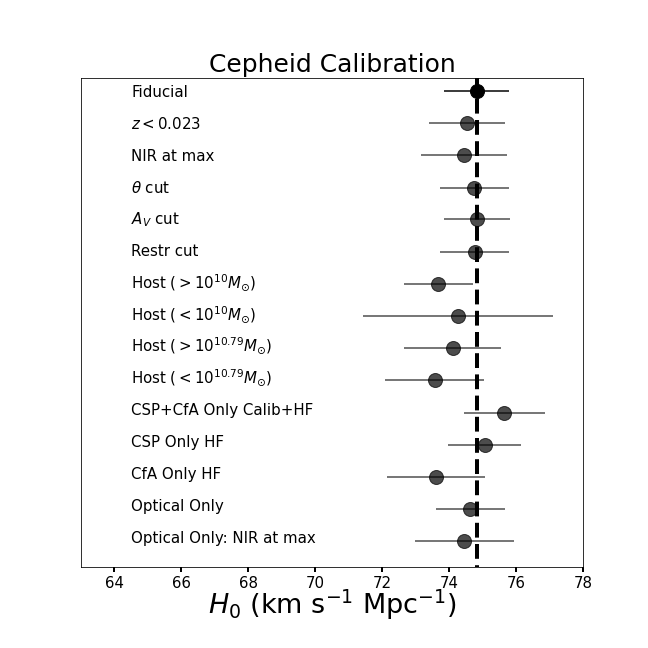}
    \includegraphics[width=.48\textwidth]{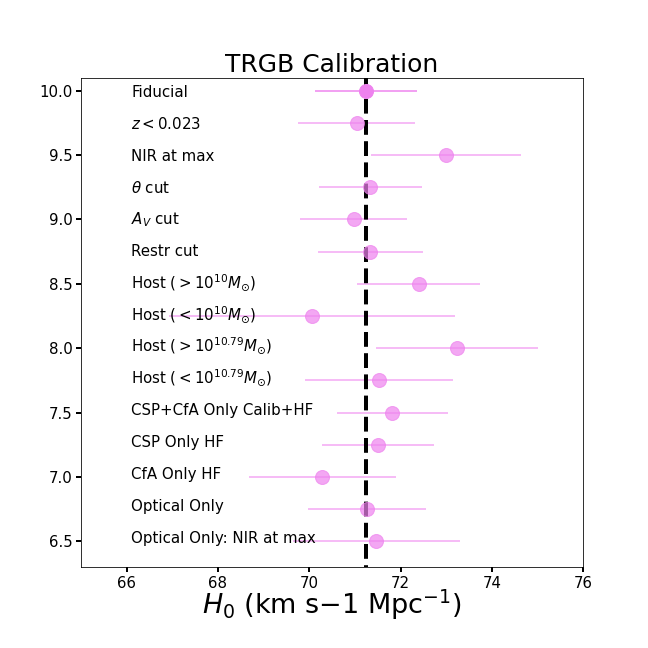}
    \caption{$H_0$ values for all the sample selection cases tried in this study for the Cepheid (left) and TRGB (right) calibration of the SN~Ia luminosity. We note that in both calibrations the largest difference from the fiducial case is for the NIR at max subsample and the subsample with SNe~Ia in low stellar mass host galaxies. The fiducial value for the Cepheid calibration is 74.82 \kmsmpc and for the TRGB is 70.92 \kmsmpc.}
    \label{fig:summary_altCase}
\end{figure*}

We then correct the redshifts for peculiar velocities using the 2M++ flow model \citep{Carrick2015}. 
We compare the distribution of the inferred $V$-band absorption from the host galaxy dust ($A_V$) and the decline rate parameters, $\theta$, between the calibrator and Hubble flow samples in Figure~\ref{fig:av_theta_dist}.  We note that the low-reddening distribution inferred for the calibrator samples from the BayeSN inference is consistent with other methods for inferring extinction for the calibrator sample \citep[e.g.][]{riess2021}. We note, however, that there are no heavily extinguished ($A_V > 1.5$) SNe in either our calibrator or Hubble flow sample. Similarly we find a larger fraction of the Hubble flow sample have $\theta > 0$ (i.e. faster declining lightcurves) compared to the calibrator sample. We test what the impact of difference in the $A_V$  and $\theta$ distributions is on the final cosmological parameters. 
\input{tables/h0_fitCases_ceph}

To infer cosmological parameters from the combination of the calibrator and Hubble flow SNe, we express the luminosity distance as a Taylor expansion in terms of time derivatives of the scale factor. First, we define the dimensionless luminosity distance as 
\begin{equation}
\begin{split}
\tilde{d}_L(z) = {z}{\left[1+\frac{(1-q_0)z}{2} - \frac{(1 - q_0 - 3
q^2_0 + j_0)z^2}{6} + \mathcal{O}(z^3)\right]}
\end{split}
\label{eq:red_lumdist}
\end{equation}
where $z$ refers to the ``true" cosmological redshift.  The luminosity distance is then $d_L(z) = c \, \tilde{d}_L(z) / H_0$ and the distance modulus is $\mu(z; H_0) = 25 + 5 \log_{10}(d_L(z) \, \text{Mpc}^{-1})$. 
We perform the standard analysis with  fixed $q_0 = -0.55$ and $j_0 = 1$. 
In the absence of errors, the intercept of the ridge line of the Hubble diagram is defined as 
\begin{equation}
\begin{split}
a = \log_{10} (c\, \tilde{d}_L(z)) - 0.2 \times \mu_{\rm SN}.
\end{split}
\label{eq:intercept}
\end{equation}
Therefore, $H_0$ can be written in terms of the absolute magnitude offset and the intercept of the Hubble diagram as
\begin{equation}
\log_{10} H_0 = \frac{\Delta M + 5a + 25}{5},
\label{eq:h0}
\end{equation}
where $\Delta M$ is the offset  from the reference absolute magnitude assumed in the SN Ia model and is constrained by the calibrator sample. 

For the calibrator samples, we define $\Delta \hat{\mu}_{\text{Cal},k} = \hat{\mu}_{\text{SN}, k} - \hat{\mu}_{\text{Cal}, k}$ where $\hat{\mu}_{\text{SN}, k}$ is the SN~Ia photometric distance estimate and $\hat{\mu}_{\text{Cal}, k}$ is the independent calibrator distance  estimate (either from Cepheids or the TRGB) to the $k$th host galaxy. Its variance is given by
\begin{equation}
\sigma_{{\rm Cal},k}^2 = \sigma_{{\rm SN, fit},k}^2 + \sigma_{{\rm Cal,}k}^2 +
\sigma_{\rm int}^2.
\label{eq:calibunc}
\end{equation}

For the Hubble flow rung, we define $\Delta \hat{\mu}_{\text{HF},i} = \hat{\mu}_{\text{SN}, i} -  \mu(\hat{z}_i; H_0^{\rm ref})$ where $H_0^{\rm ref}$ is a fixed  reference value and $\hat{z}_i$ is the inferred redshift for the $i$th Hubble flow SN~Ia  host galaxy. The final posterior distributions do not depend on the choice of $H_0^{\rm ref}$. 
The variance of $\Delta \mu_{{\rm HF},i}$ is given by
\begin{equation}
\sigma_{{\rm HF},i}^2 = \sigma_{{\rm SN, fit},i}^2  +
\sigma_{{\rm pec,mag},i}^2 + \sigma_{\rm int}^2.
\end{equation}
where
\begin{equation}
 \sigma_{{\rm pec,mag}, i} \approx \frac{5}{\ln 10}\frac{\sigma_{\rm pec}}{cz_i}
\label{eq:pecv}
\end{equation}
is the magnitude uncertainty due to the peculiar velocity errors.
We assume a peculiar velocity uncertainty $\sigma_{\rm pec} = 250\, \text{km s}^{-1}$ \citep[e.g.][]{riess2021}.

We therefore define the likelihood as the product of the calibrator and Hubble flow likelihoods
\begin{equation}
\mathcal{L}(H_0, \Delta M, \sigma_{\rm int}) = \mathcal{L_{\rm Cal}} \times \mathcal{L_{\rm HF}},
\end{equation}
 where the calibrator likelihood is
\begin{equation}
\mathcal{L}_{\rm Cal} = \prod_{k=1}^{N_{\rm Cal}} \mathcal{N}(\Delta \hat{\mu}_{\text{Cal},k} |\, \Delta M, \sigma_{\rm Cal, k}^2),
\end{equation}
where $N_{\rm Cal}$ is the total number of calibrator host galaxies.
 and the Hubble flow likelihood is
\begin{equation}
\begin{split}
\mathcal{L}_{\text{HF}} &= \prod_{i=1}^{N_{\rm HF}} \mathcal{N}(\hat{\mu}_{\text{SN}, i} - \mu(\hat{z_i}; H_0)| \, \Delta M, \sigma_{\text{HF}, i}^2) \\
&= \prod_{i=1}^{N_{\rm HF}} \mathcal{N}( \Delta \hat{\mu}_{\text{HF},i} | \, \Delta M - 5 \log_{10}(H_0/H_0^{\rm ref}), \sigma_{\text{HF}, i}^2)
\end{split}
\end{equation}
 where $N_{\rm HF}$ is the total number of Hubble flow host galaxies. Here we have used the identity $\mu(\hat{z}_{i}; H_0) = \mu(\hat{z_i}; H_0^{\rm ref}) - 5 \log_{10} (H_0 / H_0^{\rm ref})$ so that $\Delta \hat{\mu}_{\text{HF},i}$ can be computed independently of $H_0$, while the overall likelihood is still independent of $H_0^{\rm ref}$. Hence, the log likelihood we use for our analysis is,
\begin{equation}
\begin{split}
\ln \mathcal{L}(H_0, \Delta M, \sigma_{\rm int}) = &-\frac{1}{2} \sum_k \ln(2 \pi \sigma_{{\rm Cal},k}^2) + \frac{(\Delta \hat{\mu}_{{\rm Cal},k} - \Delta M)^2}{\sigma_{{\rm Cal},k}^2}\\
&-\frac{1}{2} \sum_i \ln(2 \pi \sigma_{{\rm HF},i}^2) + \frac{(\Delta \hat{\mu}_{{\rm HF},i} - \Delta M)^2}{\sigma_{{\rm HF},i}^2}, 
\end{split}
\end{equation}

The parameters $H_0$, $\Delta M$ and $\sigma_{\rm int}$  are inferred using an affine-invariant Markov Chain Monte Carlo  as implemented by the \texttt{emcee} software \citep{Foreman-Mackey2013} to sample the posterior distribution. We use 200 walkers and 20000 samples per walker with a ``burn-in" of 1000 samples. For our parameter priors, we use a uniform prior between 50 and 100 \kmsmpc for $H_0$, between $-2$ and $2$ for $\Delta M$, and between 0 and 1 for $\sigma_{\rm int}$. We infer $a$ from each MCMC sample given $H_0$ and $\Delta M$, using eq~\ref{eq:h0} and for convenience we report it as -5$a$. 
The resulting values are plotted in figure~\ref{fig:contour_H0}. We can see that, as expected, the value of $H_0$ is degenerate with $\Delta M$. We find, for the fiducial case $H_0 = 74.82 \pm 0.97\,{\rm km}\, {\rm s}^{-1}\, {\rm Mpc}^{-1}$, $\Delta M = 0.03 \pm 0.023$ mag. 
\input{tables/h0_fitCases_trgb}

Similar to the method above, we fit the calibrator sample from the Carnegie Chicago Hubble program (CCHP), a total of 18 calibrator SNe~Ia in 15 host galaxies \citep{Freedman2019}. A summary of the SN fit parameters and the TRGB distances is presented in Table~\ref{tab:calib_trgb}. 
Inferring the parameters from this calibrator sample, in combination with the same Hubble flow sample as used above, we get $H_0 = 70.92 \pm 1.14$ \kmsmpc and $\Delta M = -0.078 \pm 0.03$ mag. Similar to the Cepheid calibration, we test the impact of alternate sample selection criteria. The summary of the constraints is presented in figure~\ref{fig:summary_altCase} and table~\ref{tab:h0_trgb}. 

For the fiducial case, the scatter is $\sigma_{\rm int} = 0.091 \pm 0.014$ mag and $0.087 \pm 0.014$ mag for the Cepheid and TRGB calibration respectively. A single intrinsic scatter model can overlook systematic uncertainties. As in previous work, we test for the consistency of the dispersion in the calibrator and Hubble flow samples. We fit for two separate intrinsic scatters and find  $\sigma_{\rm int, Cal} = 0.117^{+0.021}_{-0.028}$ and $\sigma_{\rm int, HF} = 0.067^{+0.039}_{-0.025}$ for the Cepheid calibration and  $\sigma_{\rm int, Cal} = 0.123^{+0.028}_{-0.034}$ and $\sigma_{\rm int, HF} = 0.067^{+0.039}_{-0.026}$ for the TRGB calibration. For both methods, while the calibrator sample has higher scatter, both samples have consistent values at the $\sim 1 \sigma$ level. When comparing to the single scatter model, for the Cepheid calibration, we find the central value of $H_0$ only changes by 0.1 \kmsmpc with a slightly larger uncertainty giving $H_0 = 74.72 \pm 1.055$ \kmsmpc and the peak magnitude offset is $\Delta M = 0.028 \pm 0.027$ mag. For the TRGB calibration, the value of $H_0$ only changes by 0.05 \kmsmpc and the uncertainty increases to give $H_0 = 71.37 \pm 1.33$ \kmsmpc, while the peak magnitude offset is $\Delta M = -0.071 \pm 0.038$ mag. Hence, we do not find the inconsistency in the intrinsic scatter as seen previously with compilations of local SNe~Ia \citep[e.g.][]{Dhawan2018}.

\subsection{Sample selection}
\label{ssec:samp_select}
Similar to previous studies on estimating $H_0$ \citep[e.g.][]{Dhawan2018,Burns2018, riess2021}, we test the impact of changing the fiducial sample on the final value of $H_0$. We describe the individual criteria here

\begin{figure}
    \centering
    \includegraphics[width=.5\textwidth]{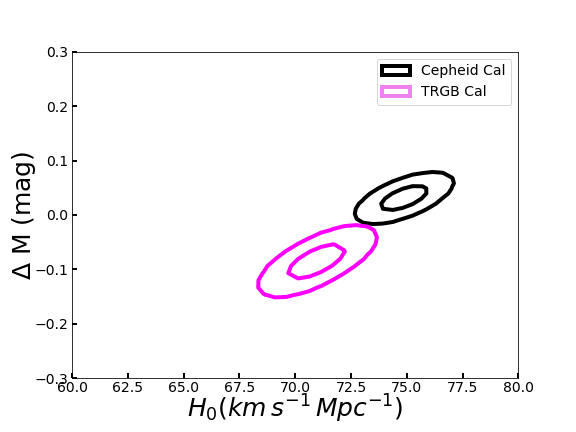}
    \caption{Comparison of the $H_0$-$M$ contours for the distance ladder calibrated to the Cepheid variables (black) and the tip of the red giant branch method (violet). }
    \label{fig:trgb_cepheid_comp}
\end{figure}
\emph{$z$-cut}: Alternate lower limit for the definition of the Hubble flow, with $z \geq 0.023$. This is an important test of systematics from peculiar velocity uncertainties on the final $H_0$ inference. While this cut only includes Hubble flow SNe~Ia where the uncertainty from peculiar motions is $\sim 4\%$ or lower, it sizably reduces the Hubble flow sample. 

\emph{NIR-at-Max-cut}: We follow the ``NIR-at-max" cut from \citet{Avelino2019}. In this case, we only use the SNe that have at least one NIR observations 2.5 days or more before time of $B$-maximum,  Since SNe~Ia are closest to standard candles near peak in the NIR \citep[e.g.][]{krisciunas2004,phillips2012,Avelino2019}, this subsample presents the estimate of $H_0$ from the best observed calibrator and Hubble flow objects. This subsample is most suited to compare the impact of adding the NIR to the data in the optical filters and hence, evaluating the improvement due to consistently modelling the optical and NIR data.

\emph{Restric-Cut}: This is most restrictive cut on the  dust $A_V$ and the lightcurve shape $\theta$ parameters to make the calibrator and Hubble flow samples as similar as possible. We restrict the Hubble flow sample to only those SNe within the range of $A_V$ and $\theta$ values  estimated for the calibrator sample (see Figure~\ref{fig:av_theta_dist}). Since current calibrator and Hubble flow SNe~Ia are derived from several heterogeneous surveys, the aim of this cut to mitigate the impact of different lightcurve shape and reddening distributions due to, e.g., individual survey selection effects. 

\emph{Survey Subsamples}: Since a large part of the Hubble flow sample is derived from either the CSP or the CfA supernova campaigns, in this cut we test the impact of using only CSP or CfA SNe for the Hubble flow rung.  We also test the shift in the value when using only SNe~Ia from CSP or CfA in both the calibrator and Hubble flow rungs.

\emph{Host galaxy mass}: We test the impact of the host galaxy mass on the final estimate of $H_0$. Since there are possibilities of differing host galaxy properties between the calibrator and Hubble flow SNe~Ia, e.g. because Cepheids are only found in young, star-forming galaxies, we test the impact of homogenizing the host galaxy property distribution. For this, we split the sample based on  $\log_{10}(M_{\rm host}/M_{\odot}) \le, > 10$  into the ``low-mass" and ``high-mass" subsamples, respectively. We also check the impact of the host mass split at the median mass, which we find to be 10$^{10.79}$ M$_{\odot}$ for our sample. This value is comparatively higher than the fiducial value of 10$^{10} M_{\odot}$. For these tests, we only use the subsample of SNe~Ia where host galaxy masses are available in the literature. For the Cepheid calibrator sample, this is the case for all 37 host galaxies, whereas there are 13 TRGB host galaxies with available mass estimates and a total of 55 Hubble flow SN~Ia host galaxies.  As not all Hubble flow host galaxies, and in the case of the TRGB calibration, not all calibrator galaxies have a stellar mass estimate, the $H_0$ inference from the host mass subsample cuts are not symmetrically distributed around the fiducial value, which uses the full sample.

\emph{Optical-Only}: These subsamples use only the optical ($\lambda < 1\,\mu$m) lightcurves for inferring the photometric distance from BayeSN. Here, we test the results for both the fiducial sample of calibrators and the subsample defined above as ``NIR at max".

 The results from the different cases are summarised in Tables~\ref{tab:calib_ceph} and~\ref{tab:calib_trgb}.
We note that out of the above cases, the largest shifts for the Cepheid calibration are for the subsample with only NIR-at-max and the low-mass sample defined with respect to the median host galaxy mass. For the TRGB calibration, the largest shifts are for the high mass subsample and the NIR at max subsamples. In both cases we see shifts of $\sim 0.7 -0.8$ \kmsmpc when using only the CSP Hubble flow sample and $\sim 1.7 - 1.8$ \kmsmpc when using only the CfA Hubble flow sample,  the latter of which is the lowest $H_0$ value in both calibration cases.
 We note that for the Cepheid calibration, there is a larger fraction of alternate cases that is below the fiducial value whereas for the TRGB case a larger fraction is above the fiducial value. We note, however, that in all these cases, the perturbations are consistent with the fiducial value, within errors, and we, therefore, conclude that these is likely to be a statistical fluctuations.

\subsection{Systematic Uncertainties}
In this section we quantify the systematic uncertainty on the inferred value of $H_0$ for both the cases with the Cepheid and TRGB calibrations. 
\paragraph*{Cepheid Calibration}: We use the sample of Cepheid calibrators from the R22 dataset. The Cepheid systematics for this dataset are at the 0.7$\%$ \citep[Table 7 in ][]{riess2021}. We add the contribution from the standard deviation our analysis variants in Section~\ref{ssec:samp_select}. The total systematic error is 0.84 \kmsmpc. 

\paragraph*{TRGB calibration}: 
For the TRGB calibrated distance ladder, we take the systematic error from the absolute calibration of the TRGB magnitude as reported in \citet{Freedman2021}. An error in the $M_{\rm I, TRGB}$ of 0.038 mag corresponds to a $\sigma(H_0)$ of 1.33 \kmsmpc. Similar to the case for the Cepheid calibrated distance ladder, we add the standard deviation of our analysis variants, hence, the total systematic uncertainty is 1.49 \kmsmpc. 

We note that taking the standard deviation of the analysis variants is a conservative upper limit on the systematic uncertainty. 

\section{Discussion and Conclusions}
\label{sec:disc}
We have presented a consistent inference of $H_0$ from modelling the optical to near infrared lightcurves of SNe~Ia. We find $H_0 = 74.82 \pm 0.97  \pm 0.84$ \kmsmpc\, for the Cepheid calibration and $H_0 = 70.92 \pm 1.14 \pm 1.49   $ \kmsmpc\, for the TRGB calibration.= These value of $H_0$ and $\sim 1 - 1.5$ \kmsmpc higher than the cases estimated in the literature, using optical only data for SNe~Ia \citep{riess2021,Freedman2021}. This is also seen in other studies incorporating the NIR SN~Ia data \citep[e.g.][]{galbany2022}.  We note that while the Cepheid calibration measurement is in a significant (4.9$\sigma$) tension with the early universe inference from \citet{2020A&A...641A...6P}, the TRGB measurement has a greater consistency with the early universe inference (1.8$\sigma$ tension). The level of tension for each of the two probes is very similar to reported degrees of tension in the literature \citep[e.g.,][]{Freedman2021,riess2021}.

We test the impact of various analysis assumptions for both the calibrator and Hubble flow samples for both the Cepheid and TRGB calibrations. We find that for the Cepheid calibration, the alternate assumptions do not change the central value of $H_0$ significantly. However, in the cases of low host galaxy mass subsample the uncertainty is more than twice larger which is due mostly to a smaller number of SNe in the sample. For the $A_V$ and $\theta$ cuts as well as the Restrictive cut applied in the sample, the TRGB calibrators have a more similar distribution to the Hubble flow compared to the Cepheid calibrators.

We test the impact of simultaneous modelling of the optical and the NIR by fitting only the optical filters for both the calibrator and the Hubble flow sample. For the complete sample, we do not find any significant shift in the value of $H_0$ from the optical only, with respect to the fiducial case. Moreover, for the complete sample, the improvement in the final $H_0$ estimate is marginal. We attribute this to the the dominant source of uncertainty arising from the small number of calibrator SNe~Ia. Additionally since both the calibrator and Hubble flow SNe~Ia have densely sampled lightcurves, we compare the $H_0$ estimate and uncertainty from only the NIR-at-max subsample. The $H_0$ constraint from this subsample is up to $\sim$15\% more precise than the constraint from only fitting the optical lightcurves for the same subsample. This is  a promising sign for using consistent modelling of SNe~Ia lightcurves from the optical to the NIR, even for high-$z$ SN cosmology studies, e.g. with the RAISIN survey \citep{Jones2022:raisin}. 

 There have been debates in the literature as to whether effects of host galaxy environment can highly bias the inferred $H_0$ \citep{Rigault2020,Jones2018}. Here we tested the impact of the environment by dividing the samples into low- and high-mass host galaxies In the first case, we divide the sample in mass bins above or below 10$^{10}$ M$_{\odot}$, as done previously in the literature \citep[e.g.][]{riess2021}. While the value of $H_0$ is shifted in the low-mass sample compared to the fiducial analyses, the uncertainty  is larger due to the small number of SNe~Ia in the subsample in the current analyses. Hence, it will be crucial to test this effect with SN~Ia samples (both in the calibrator and Hubble flow) derived from untargeted surveys such as ZTF \citep{Dhawan2022a} and YSE \citep{Jones2021}. Moreover, we note that the high host mass subsample has a lower $\sigma_{\rm int}$. The SN~Ia datasets from untargeted surveys will be critical to understand what the origin of such a low scatter is.

Aspects of our analysis can be improved in terms of both statistical and systematic uncertainties. Our statistical uncertainties would be improved with a larger number of SNe in both the calibrator and Hubble flow datasets. In the near future, observatories like the James Webb Space Telescope have the capabilities to observe host galaxies of SNe~Ia in the local volume out to $\sim 80$\, Mpc, significantly larger than the farthest calibrator galaxy in the current sample at $\sim 40$\, Mpc \citep[e.g.][]{riess2021}. This can increase the calibrator sample from a few tens of SNe~Ia to $> 100$ SNe~Ia, when using methods like the TRGB which are ``single-shot" and do not require multiple visits per galaxy to get an accurate distance. Having such an untargeted sample will also be important for testing the impact of host galaxy properties on the inferred $H_0$, since the current Hubble flow sample has small number statistics in the low mass bin.
We note, additionally, that in this analysis we use the most updated calibration of the Cepheid and TRGB methods. However, there are alternate assumptions in the more nearby rungs of the Cepheid \citep{Mortsell2021a,Mortsell2021b} or TRGB \citep{Anand2022} distance ladder,  e.g. a 0.05 mag brighter (fainter) shift in the absolute magnitude would propagate throughout the distance ladder measurement as a shift in $H_0$ of $\sim$ 1.6 \kmsmpc lower (higher), \citep[e.g., see][for a summary of the absolute calibration of the TRGB]{Li2022}. 
Moreover, our current analysis uses peculiar velocity corrections from the 2M++ flow model as described in \citet{Carrick2015}. Recent improvements in the corrections can further help reduce the uncertainty in the inferred $H_0$ \citep{Peterson2021,Kenworthy2022}.

There are possibilities to reduce systematic uncertainties in the future. Improving the BayeSN model by retraining using an updated set of NIR spectra \citep[e.g.][]{Lu2022} can further reduce uncertainties in the inferred distances from the SED model. Since we required a dataset with both optical and NIR coverage, the median redshift of the sample is typically lower than that of samples in the literature. A sample with higher median $z$ would be less prone to systematics from peculiar velocities. A higher median redshift is one of salient features of SN~Ia programs optimised for the NIR, e.g. SIRAH \citep{Jha2019}. In addition, for our work, we use a fixed global value of the dust $R_V$ for all objects. However, we can use  joint optical and NIR data to estimate $R_V$ \citep{Krisciunas2007,Mandel2011,Burns2014}. Such optical-NIR constraints will be important to probe potential differences in intrinsic or dust properties in various SN~Ia host galaxy environments  \citep[e.g.,][]{uddin2020, BS20,Johansson2021,Thorp2021,Thorp2022} and test the impact of possible differences on the inferred value of $H_0$. Moreover, some of the surveys that observed SNe in our sample were targeted to particular galaxy types. Modelling a uniform distance ladder with both calibrator and Hubble flow SNe~Ia discovered and followed-up with the same survey could reduce this source of systematic error \citep[e.g.][]{Dhawan2022b}. Our results show potential pathways to reduce uncertainties in the measurement of a key cosmological parameter, $H_0$, with future large samples of SNe~Ia.

\section*{acknowledgements}
We thank Mat Smith for useful discussions on bandpasses. SD acknowledges support from the European Union's Horizon 2020 research and innovation programme Marie Skłodowska-Curie Individual Fellowship (grant agreement No. 890695), and a Junior Research Fellowship at Lucy Cavendish College, Cambridge. ST was supported by the Cambridge Centre for Doctoral Training in Data-Intensive Science funded by the UK Science and Technology Facilities Council (STFC). KSM acknowledges funding from the European Research Council under the European Union’s Horizon 2020 research and innovation programme (ERC Grant Agreement No. 101002652). This project has been made possible through the ASTROSTAT-II collaboration, enabled by the Horizon 2020, EU Grant Agreement No. 873089. SMW is supported by the
UK Science and Technology Facilities Council (STFC). TC acknowledges support from the 2021 Institute of Astronomy David and Bridget Jacob Summer Research Programme.

\section*{Data Availability}
All data used in this manuscript are publicly available from the sources described in \S \ref{sec:data}.

\bibliographystyle{mnras}
\bibliography{mnras_template}


\bsp	
\label{lastpage}
\end{document}

%% file: tables/calib_DM_Ceph.tex
\begin{table*}
\centering
\begin{minipage}{\textwidth}
\caption{Calibrator sample used for inferring $H_0$ with independent distances to the host galaxy derived from Cepheid variables. The SN distance and uncertainty are reported along with the inferred lightcurve shape parameter $\theta$ and the absorption in the $V$-band, $A_V$. We also report the logarithm of the host stellar mass and whether the sample is in the NIR at max subsample (see text for details). The table is truncated for formatting reasons, entire table is available online.}
\begin{tabular}{|c|c|c|c|c|c|c|c|c|c|c|}
\hline 
SN & Host &  $\mu_{\rm SN}$ & $\sigma_{\rm SN, fit}$ & $\mu_{\rm Ceph}$ & $\sigma_{\rm Ceph}$ & $\theta$ & $A_V$ & log$_{10}$ Host mass & NIR at max & Photometry Reference \footnote{[1]: \citet{Hamuy1991}, [2]: \citet{elias1981}, [3]: \citet{Leibundgut1991}, [4]: \citet{riess2005}. [5]: \citet{riess1999}, [6]: \citet{scolnic2021}} \\
\hline
1981B & NGC 4536 & 30.893 & 0.054 & 30.838 & 0.051 & 0.076 & 0.253 & 9.69 & N & [1,2] \\
1990N & NGC 4639 & 31.803 & 0.058 & 31.818 & 0.085 & -0.701 & 0.168 & 9.80 & N & [3] \\
1994ae & NGC 3370 & 32.019 & 0.059 & 32.123 & 0.052 & -0.426 & 0.308 & 10.20 & N & [4] \\
1995al & NGC 3021 & 32.234 & 0.062 & 32.475 & 0.160 & -0.956 & 0.356 & 10.30 & N & [5] \\
1997bp & NGC 4680  & 32.524 & 0.052 & 32.606 & 0.208 & 0.213 & 0.479 & 9.75 & N & [6]\\
\hline
\end{tabular}
\end{minipage}
\label{tab:calib_ceph}
\end{table*}

%% file: tables/calib_DM_TRGB.tex
\begin{table*}
\centering
\begin{minipage}{\textwidth}
\caption{Similar to Table~\ref{tab:calib_ceph}, the calibrator sample using the TRGB method.}
\begin{tabular}{|c|c|c|c|c|c|c|c|c|c|c|}
\hline
SN & Host & $\mu_{\rm SN}$ & $\sigma_{\rm SN, fit}$ & $\mu_{\rm TRGB}$ & $\sigma_{\rm TRGB}$ & $\theta$ & $A_V$ & log$_{10}$ Host mass & NIR at max & Photometry reference \footnote{[1]:\citet{Hamuy1991}, [2]:\citet{elias1981}, [3]: \citet{Tsvetkov1982}, [4]:\citet{Walker1982}, [5]: \citet{Wells1994}, [6]: \citet{Richmond1995}, [7]: \citet{riess2005}, [8]: \citet{riess1999}, [9]: \citet{Jha1999}, [10]: \citet{Krisciunas2003}, [11]: \citet{Silverman2012},[12]:\citet{Cartier2014}, [13]: \citet{Stritzinger2010}, [14]: \citet{Stritzinger2011}, [15]: \citet{Gall2018} [16]: \citet{Schweizer2008}, [17]: \citet{Richmond2012}, [18]: \citet{Matheson2012},   [19]: \citet{Marion2016}, [20]: \citet{Contreras2018} }\\
\hline
SN1980N & NGC 1316 & 31.406 & 0.055 & 31.46 & 0.04 & 0.131 & 0.177 & 11.57 & N & [1,2]\\
SN1981B & NGC 4536 & 30.893 & 0.054 & 30.96 & 0.05 & 0.076 & 0.253 & 10.47 & N & [3,2] \\
SN1981D & NGC 1316 & 31.133 & 0.069 & 31.46 & 0.04 & 0.862 & 0.391 & 11.57 & N & [4,2]\\
SN1989B & NGC 3627 & 30.138 & 0.060 & 30.22 & 0.04 & -0.459 & 1.049 & $\ldots$ & N & [5] \\
SN1994D & NGC 4526 & 30.778 & 0.035 & 31.00 & 0.07 & 1.726 & 0.074 & 11.00 & N & [6]\\
SN1994ae & NGC 3370 & 32.019 & 0.059 & 32.27 & 0.05 & -0.426 & 0.308 & 9.69 & N & [7]\\
SN1995al & NGC 3021 & 32.234 & 0.062 & 32.22 & 0.05 & -0.957 & 0.357 & 9.87 & N & [8] \\
SN1998bu & NGC 3368  & 29.973 & 0.037 & 30.31 & 0.04 & -0.125 & 1.006 & $\ldots$ & Y & [9] \\
SN2001el & NGC 1448 & 31.329 & 0.034 & 31.32 & 0.06 & -0.303 & 0.589 & 10.69 & Y & [10]\\
SN2002fk & NGC 1309 & 32.434 & 0.035 & 32.50 & 0.07 & -0.386 & 0.102 & 9.94 & Y & [11, 12] \\
SN2006dd & NGC 1316 & 31.247 & 0.040 & 31.46 & 0.04 & 0.523 & 0.302 & 11.57 & Y & [13] \\
SN2007af & NGC 5584 & 31.938 & 0.035 & 31.82 & 0.10 & 0.302 & 0.323 & 9.49 & Y & [14] \\
SN2007on & NGC 1404 & 31.706 & 0.032 & 31.42 & 0.05 & 2.458 & 0.240 & 11.17 & N & [15] \\
SN2007sr & NGC 4038 & 31.621 & 0.040 & 31.68 & 0.05 & -0.427 & 0.330 & 10.05 & N & [16]  \\
SN2011fe & M101 & 29.054 & 0.043 & 29.08 & 0.04 & -0.170 & 0.184 & 12.20 & Y & [17, 18] \\
SN2011iv & NGC 1404 & 31.056&  0.034 & 31.42 & 0.05 & 1.756 & 0.357 & 11.17 & Y & [15] \\
SN2012cg & NGC 4424 & 31.148 & 0.043 & 31.00 & 0.06 & -0.797 & 0.182 & 8.47 & Y & [19] \\
SN2012fr & NGC 1365 & 31.365 & 0.025 & 31.36 & 0.05 & -1.271 & 0.015 & 6.74 & Y & [20] \\
\hline
\end{tabular}
\end{minipage}
\label{tab:calib_trgb}
\end{table*}

%% file: tables/hubbleflow.tex
\begin{table*}
\caption{The Hubble flow sample used in our inference. CMB frame redshifts with peculiar velocity corrections applied (referred to here as Hubble diagram redshifts) are reported along with the fitted distance from BayeSN and the lightcurve shape and absorption parameter inferred from the model fit. We emphasise that this distance, $\mu_{\rm SN}$ is inferred from the same scaling as for the calibrator sample in Tables~\ref{tab:calib_ceph} and Table~\ref{tab:calib_trgb}, hence, does not impact the final estimate of $H_0$. The table here is truncated for formatting reasons and is available in its entirety online.}
\begin{tabular}{|c|c|c|c|c|c|c|c|}
\hline
SN & $z_{\rm HD}$ & $\mu_{\rm SN}$ & $\sigma_{\rm SN, fit}$   & $\theta$ & $A_V$ & $\log_{10}$ Host mass & NIR at max\\
\hline
1999ee & 0.01122 & 33.175 & 0.035 & -0.291 & 0.712 & $\ldots$ & Y \\
1999ek & 0.017821 & 34.196 & 0.040 & 0.143 & 0.710 & $\ldots$ & Y \\
2000bh & 0.024195 & 34.889 & 0.044 & -0.129 & 0.189 & $\ldots$ & N \\
2000ca & 0.02391 & 34.987 & 0.034 & 0.189 & 0.039 & $\ldots$ & Y \\
\hline
\end{tabular}
\label{tab:hubbleflow}
\end{table*}

%% file: tables/h0_fitCases_ceph.tex
\begin{table*}
    \centering
        \caption{Inferred value of $H_0$ and the absolute luminosity offset for the fiducial case and alternate case scenarios for the Cepheid calibrated distance ladder. We also report the intrinsic scatter term and the number of SNe in each of the calibrator and Hubble flow samples in every case.}

\begin{tabular}{|c|c|c|c|c|c|c|}
\hline
    Case & $H_0$    &   $\Delta M$ &  $-5a$ \footnote{computed using equation~\ref{eq:h0}} &  $\sigma_{\rm int}$ & $N_{\rm calib}$ & $N_{\rm HF} $ \\
    & km\,${\rm s}^{-1}$\,${\rm Mpc} ^{-1}$ & mag & mag & mag \\
    \hline
Fiducial & 74.823 $\pm$ 0.973 & 0.030 $\pm$ 0.023 & 15.660 $\pm$ 0.016 & 0.090$^{+0.013}_{-0.013}$ &37 & 67  \\ 
$z < 0.023$ & 74.532 $\pm$ 1.135 & 0.030 $\pm$ 0.024 & 15.668 $\pm$ 0.022 & 0.097$^{+0.016}_{-0.015}$ &37 & 31  \\ 
NIR at max & 74.440 $\pm$ 1.286 & 0.008 $\pm$ 0.032 & 15.649 $\pm$ 0.020 & 0.079$^{+0.018}_{-0.016}$ & 15 & 40  \\ 
$\theta$ cut & 74.751 $\pm$ 1.021 & 0.030 $\pm$ 0.024 & 15.662 $\pm$ 0.018 & 0.094$^{+0.015}_{-0.013}$ &37 & 58  \\ 
$A_V$ cut & 74.821 $\pm$ 0.992 & 0.030 $\pm$ 0.023 & 15.660 $\pm$ 0.017 & 0.090$^{+0.014}_{-0.013}$ &37 & 65  \\ 
Restr cut & 74.756 $\pm$ 1.035 & 0.030 $\pm$ 0.024 & 15.662 $\pm$ 0.018 & 0.095$^{+0.015}_{-0.014}$ &37 & 56  \\ 
Host ($> 10^{10} M_{\odot}$) & 73.678 $\pm$ 1.023 & -0.012 $\pm$ 0.025 & 15.651 $\pm$ 0.016 & 0.068$^{+0.015}_{-0.015}$ &25 & 49  \\ 
Host ($< 10^{10} M_{\odot}$) & 74.259 $\pm$ 2.842 & 0.119 $\pm$ 0.053 & 15.765 $\pm$ 0.064 & 0.133$^{+0.043}_{-0.035}$ & 12 & 7  \\ 
Host ($> 10^{10.79} M_{\odot}$) & 74.102 $\pm$ 1.441 & -0.015 $\pm$ 0.037 & 15.636 $\pm$ 0.021 & 0.070$^{+0.022}_{-0.019}$ &13 & 32  \\ 
Host ($< 10^{10.79} M_{\odot}$) & 73.561 $\pm$ 1.484 & 0.052 $\pm$ 0.031 & 15.718 $\pm$ 0.031 & 0.104$^{+0.022}_{-0.019}$ &24 & 22  \\ 
 CSP+CfA only calib+HF &  75.641 $\pm$ 1.214 &  0.059 $\pm$ 0.031 &  15.666 $\pm$ 0.016 &  0.077$^{+0.017}_{-0.015}$ &  19 &  58  \\
CSP Only HF & 75.056 $\pm$ 1.091 & 0.030 $\pm$ 0.024 & 15.653 $\pm$ 0.021 & 0.092$^{+0.016}_{-0.015}$ &37 & 39  \\ 
CfA Only HF & 73.591 $\pm$ 1.462 & 0.029 $\pm$ 0.026 & 15.695 $\pm$ 0.035 & 0.109$^{+0.021}_{-0.019}$ &37 & 19  \\ 
Opt-Only & 74.630 $\pm$ 1.021 & 0.025 $\pm$ 0.022 & 15.660 $\pm$ 0.020 & 0.088$^{+0.016}_{-0.015}$ &37 & 67  \\ 
Opt-Only: NIR at Max & 74.452 $\pm$ 1.490 & 0.007 $\pm$ 0.034 & 15.647 $\pm$ 0.026 & 0.087$^{+0.023}_{-0.020}$ &15 & 40  \\ 
 \hline
    \end{tabular}
    \label{tab:h0}
\end{table*}

%% file: tables/h0_fitCases_trgb.tex
\begin{table*}
    \centering
        \caption{Inferred value of $H_0$ and the absolute luminosity offset for the fiducial case and alternate case scenarios for the TRGB calibrated distance ladder. We also report the intrinsic scatter term and the number of SNe in each of the calibrator and Hubble flow samples in every case.}
\begin{tabular}{|c|c|c|c|c|c|c|}
\hline
    Case & $H_0$    &   $\Delta M$ & $-5a$~\footnote{computed using equation~\ref{eq:h0}} & $\sigma_{\rm int}$ & $N_{\rm calib}$ & $N_{\rm HF} $ \\
    & km\,${\rm s}^{-1}$\,${\rm Mpc} ^{-1}$ & mag & mag & mag  \\
    \hline
Fiducial & 70.918 $\pm$ 1.149 & -0.086 $\pm$ 0.030 & 15.660 $\pm$ 0.018 & 0.085$^{+0.015}_{-0.012}$ &15 & 67  \\ 
$z < 0.023$ & 70.710 $\pm$ 1.389 & -0.085 $\pm$ 0.033 & 15.668 $\pm$ 0.026 & 0.092$^{+0.017}_{-0.015}$ &15 & 31  \\ 
NIR at max & 71.059 $\pm$ 1.557 & -0.094 $\pm$ 0.042 & 15.648 $\pm$ 0.023 & 0.090$^{+0.018}_{-0.015}$ & 9 & 40  \\ 
$\theta$ cut & 70.962 $\pm$ 1.185 & -0.086 $\pm$ 0.031 & 15.659 $\pm$ 0.019 & 0.086$^{+0.015}_{-0.013}$ &15 & 62  \\ 
$A_V$ cut & 70.972 $\pm$ 1.171 & -0.085 $\pm$ 0.031 & 15.659 $\pm$ 0.018 & 0.108$^{+0.014}_{-0.012}$ &15 & 67  \\ 
Restr cut & 70.984 $\pm$ 1.173 & -0.085 $\pm$ 0.031 & 15.659 $\pm$ 0.019 & 0.086$^{+0.014}_{-0.013}$ &15 & 62  \\ 
Host ($> 10^{10} M_{\odot}$) & 71.225 $\pm$ 1.454 & -0.086 $\pm$ 0.040 & 15.651 $\pm$ 0.020 & 0.044$^{+0.021}_{-0.019}$ &7 & 49  \\ 
Host ($< 10^{10} M_{\odot}$) & 70.111 $\pm$ 3.170 & -0.006 $\pm$ 0.071 & 15.764 $\pm$ 0.069 & 0.137$^{+0.053}_{-0.041}$ & 6 & 7  \\ 
Host ($> 10^{10.79} M_{\odot}$) & 71.289 $\pm$ 1.959 & -0.095 $\pm$ 0.052 & 15.634 $\pm$ 0.029 & 0.040$^{+0.020}_{-0.028}$ &5 & 29  \\ 
Host ($< 10^{10.79} M_{\odot}$) & 71.608 $\pm$ 1.618 & -0.023 $\pm$ 0.042 & 15.702 $\pm$ 0.025 & 0.088$^{+0.022}_{-0.022}$ & 8 & 26  \\ 
 CSP+CfA only calib+HF & 71.822 $\pm$ 1.536 &  -0.052 $\pm$ 0.043 &  15.666 $\pm$ 0.017 &  0.088$^{+0.016}_{-0.014}$ &  9 &  58  \\
CSP Only HF & 71.188 $\pm$ 1.326 & -0.085 $\pm$ 0.032 & 15.653 $\pm$ 0.024 & 0.115$^{+0.015}_{-0.016}$ &15 & 39  \\ 
CfA Only HF & 69.792 $\pm$ 1.793 & -0.085 $\pm$ 0.038 & 15.696 $\pm$ 0.041 & 0.123$^{+0.022}_{-0.019}$ &15 & 19  \\ 
Opt-Only & 71.320 $\pm$ 1.248 & -0.057 $\pm$ 0.033 & 15.679 $\pm$ 0.018 & 0.086$^{+0.017}_{-0.014}$ &15 & 67  \\ 
Opt-Only: NIR at Max & 71.510 $\pm$ 1.732 & -0.044 $\pm$ 0.046 & 15.683 $\pm$ 0.024 & 0.105$^{+0.023}_{-0.020}$ & 9 & 40  \\ 

 \hline
    \end{tabular}
    \label{tab:h0_trgb}
\end{table*}

%% file: mnras_template.bbl
\begin{thebibliography}{}
\makeatletter
\relax
\def\mn@urlcharsother{\let\do\@makeother \do\$\do\&\do\#\do\^\do\_\do\%\do\~}
\def\mn@doi{\begingroup\mn@urlcharsother \@ifnextchar [ {\mn@doi@}
  {\mn@doi@[]}}
\def\mn@doi@[#1]#2{\def\@tempa{#1}\ifx\@tempa\@empty \href
  {http://dx.doi.org/#2} {doi:#2}\else \href {http://dx.doi.org/#2} {#1}\fi
  \endgroup}
\def\mn@eprint#1#2{\mn@eprint@#1:#2::\@nil}
\def\mn@eprint@arXiv#1{\href {http://arxiv.org/abs/#1} {{\tt arXiv:#1}}}
\def\mn@eprint@dblp#1{\href {http://dblp.uni-trier.de/rec/bibtex/#1.xml}
  {dblp:#1}}
\def\mn@eprint@#1:#2:#3:#4\@nil{\def\@tempa {#1}\def\@tempb {#2}\def\@tempc
  {#3}\ifx \@tempc \@empty \let \@tempc \@tempb \let \@tempb \@tempa \fi \ifx
  \@tempb \@empty \def\@tempb {arXiv}\fi \@ifundefined
  {mn@eprint@\@tempb}{\@tempb:\@tempc}{\expandafter \expandafter \csname
  mn@eprint@\@tempb\endcsname \expandafter{\@tempc}}}

\bibitem[\protect\citeauthoryear{{Anand}, {Tully}, {Rizzi}, {Riess}  \&
  {Yuan}}{{Anand} et~al.}{2022}]{Anand2022}
{Anand} G.~S.,  {Tully} R.~B.,  {Rizzi} L.,  {Riess} A.~G.,   {Yuan} W.,  2022,
  \mn@doi [\apj] {10.3847/1538-4357/ac68df}, \href
  {https://ui.adsabs.harvard.edu/abs/2022ApJ...932...15A} {932, 15}

\bibitem[\protect\citeauthoryear{{Avelino}, {Friedman}, {Mandel}, {Jones},
  {Challis}  \& {Kirshner}}{{Avelino} et~al.}{2019}]{Avelino2019}
{Avelino} A.,  {Friedman} A.~S.,  {Mandel} K.~S.,  {Jones} D.~O.,  {Challis}
  P.~J.,   {Kirshner} R.~P.,  2019, \mn@doi [\apj] {10.3847/1538-4357/ab2a16},
  \href {https://ui.adsabs.harvard.edu/abs/2019ApJ...887..106A} {887, 106}

\bibitem[\protect\citeauthoryear{{Barone-Nugent} et~al.,}{{Barone-Nugent}
  et~al.}{2012}]{Barone-Nugent2012}
{Barone-Nugent} R.~L.,  et~al., 2012, \mn@doi [\mnras]
  {10.1111/j.1365-2966.2012.21412.x}, \href
  {https://ui.adsabs.harvard.edu/abs/2012MNRAS.425.1007B} {425, 1007}

\bibitem[\protect\citeauthoryear{{Betancourt}}{{Betancourt}}{2016}]{Betancourt16}
{Betancourt} M.,  2016, arXiv e-prints, \href
  {https://ui.adsabs.harvard.edu/abs/2016arXiv160100225B} {p. arXiv:1601.00225}

\bibitem[\protect\citeauthoryear{{Brout} \& {Scolnic}}{{Brout} \&
  {Scolnic}}{2021}]{BS20}
{Brout} D.,  {Scolnic} D.,  2021, \mn@doi [\apj] {10.3847/1538-4357/abd69b},
  \href {https://ui.adsabs.harvard.edu/abs/2021ApJ...909...26B} {909, 26}

\bibitem[\protect\citeauthoryear{{Brout} et~al.,}{{Brout}
  et~al.}{2021}]{brout2021}
{Brout} D.,  et~al., 2021, arXiv e-prints, \href
  {https://ui.adsabs.harvard.edu/abs/2021arXiv211203864B} {p. arXiv:2112.03864}

\bibitem[\protect\citeauthoryear{{Burns} et~al.,}{{Burns}
  et~al.}{2011}]{Burns2011}
{Burns} C.~R.,  et~al., 2011, \mn@doi [\aj] {10.1088/0004-6256/141/1/19}, \href
  {http://adsabs.harvard.edu/abs/2011AJ....141...19B} {141, 19}

\bibitem[\protect\citeauthoryear{{Burns} et~al.,}{{Burns}
  et~al.}{2014}]{Burns2014}
{Burns} C.~R.,  et~al., 2014, \mn@doi [\apj] {10.1088/0004-637X/789/1/32},
  \href {https://ui.adsabs.harvard.edu/abs/2014ApJ...789...32B} {789, 32}

\bibitem[\protect\citeauthoryear{{Burns} et~al.,}{{Burns}
  et~al.}{2018}]{Burns2018}
{Burns} C.~R.,  et~al., 2018, \mn@doi [\apj] {10.3847/1538-4357/aae51c}, \href
  {https://ui.adsabs.harvard.edu/abs/2018ApJ...869...56B} {869, 56}

\bibitem[\protect\citeauthoryear{{Carpenter} et~al.,}{{Carpenter}
  et~al.}{2017}]{Carpenter2017}
{Carpenter} B.,  et~al., 2017, Journal of Statistical Software, \href
  {https://ui.adsabs.harvard.edu/abs/2017JSS....76....1C} {76, 1}

\bibitem[\protect\citeauthoryear{{Carr}, {Davis}, {Scolnic}, {Said}, {Brout},
  {Peterson}  \& {Kessler}}{{Carr} et~al.}{2021}]{Carr2021}
{Carr} A.,  {Davis} T.~M.,  {Scolnic} D.,  {Said} K.,  {Brout} D.,  {Peterson}
  E.~R.,   {Kessler} R.,  2021, arXiv e-prints, \href
  {https://ui.adsabs.harvard.edu/abs/2021arXiv211201471C} {p. arXiv:2112.01471}

\bibitem[\protect\citeauthoryear{{Carrick}, {Turnbull}, {Lavaux}  \&
  {Hudson}}{{Carrick} et~al.}{2015}]{Carrick2015}
{Carrick} J.,  {Turnbull} S.~J.,  {Lavaux} G.,   {Hudson} M.~J.,  2015, \mn@doi
  [\mnras] {10.1093/mnras/stv547}, \href
  {https://ui.adsabs.harvard.edu/abs/2015MNRAS.450..317C} {450, 317}

\bibitem[\protect\citeauthoryear{{Cartier} et~al.,}{{Cartier}
  et~al.}{2014}]{Cartier2014}
{Cartier} R.,  et~al., 2014, \mn@doi [\apj] {10.1088/0004-637X/789/1/89}, \href
  {http://adsabs.harvard.edu/abs/2014ApJ...789...89C} {789, 89}

\bibitem[\protect\citeauthoryear{{Contreras} et~al.,}{{Contreras}
  et~al.}{2018}]{Contreras2018}
{Contreras} C.,  et~al., 2018, \mn@doi [\apj] {10.3847/1538-4357/aabaf8}, \href
  {https://ui.adsabs.harvard.edu/abs/2018ApJ...859...24C} {859, 24}

\bibitem[\protect\citeauthoryear{{Davis}, {Hinton}, {Howlett}  \&
  {Calcino}}{{Davis} et~al.}{2019}]{Davis2019}
{Davis} T.~M.,  {Hinton} S.~R.,  {Howlett} C.,   {Calcino} J.,  2019, \mn@doi
  [\mnras] {10.1093/mnras/stz2652}, \href
  {https://ui.adsabs.harvard.edu/abs/2019MNRAS.490.2948D} {490, 2948}

\bibitem[\protect\citeauthoryear{{Dhawan}, {Jha}  \& {Leibundgut}}{{Dhawan}
  et~al.}{2018}]{Dhawan2018}
{Dhawan} S.,  {Jha} S.~W.,   {Leibundgut} B.,  2018, \mn@doi [\aap]
  {10.1051/0004-6361/201731501}, \href
  {https://ui.adsabs.harvard.edu/abs/2018A&A...609A..72D} {609, A72}

\bibitem[\protect\citeauthoryear{{Dhawan} et~al.,}{{Dhawan}
  et~al.}{2022a}]{Dhawan2022a}
{Dhawan} S.,  et~al., 2022a, \mn@doi [\mnras] {10.1093/mnras/stab3093}, \href
  {https://ui.adsabs.harvard.edu/abs/2022MNRAS.510.2228D} {510, 2228}

\bibitem[\protect\citeauthoryear{{Dhawan} et~al.,}{{Dhawan}
  et~al.}{2022b}]{Dhawan2022b}
{Dhawan} S.,  et~al., 2022b, \mn@doi [\apj] {10.3847/1538-4357/ac7ceb}, \href
  {https://ui.adsabs.harvard.edu/abs/2022ApJ...934..185D} {934, 185}

\bibitem[\protect\citeauthoryear{{Elias}, {Frogel}, {Hackwell}  \&
  {Persson}}{{Elias} et~al.}{1981}]{elias1981}
{Elias} J.~H.,  {Frogel} J.~A.,  {Hackwell} J.~A.,   {Persson} S.~E.,  1981,
  \mn@doi [\apjl] {10.1086/183683}, \href
  {https://ui.adsabs.harvard.edu/abs/1981ApJ...251L..13E} {251, L13}

\bibitem[\protect\citeauthoryear{{Elias}, {Matthews}, {Neugebauer}  \&
  {Persson}}{{Elias} et~al.}{1985}]{Elias1985}
{Elias} J.~H.,  {Matthews} K.,  {Neugebauer} G.,   {Persson} S.~E.,  1985,
  \mn@doi [\apj] {10.1086/163456}, \href
  {https://ui.adsabs.harvard.edu/abs/1985ApJ...296..379E} {296, 379}

\bibitem[\protect\citeauthoryear{{Fitzpatrick}}{{Fitzpatrick}}{1999}]{Fitzpatrick1999}
{Fitzpatrick} E.~L.,  1999, \mn@doi [\pasp] {10.1086/316293}, \href
  {https://ui.adsabs.harvard.edu/abs/1999PASP..111...63F} {111, 63}

\bibitem[\protect\citeauthoryear{{Folatelli} et~al.,}{{Folatelli}
  et~al.}{2010}]{Folatelli2010}
{Folatelli} G.,  et~al., 2010, \mn@doi [\aj] {10.1088/0004-6256/139/1/120},
  \href {https://ui.adsabs.harvard.edu/abs/2010AJ....139..120F} {139, 120}

\bibitem[\protect\citeauthoryear{{Foreman-Mackey}, {Hogg}, {Lang}  \&
  {Goodman}}{{Foreman-Mackey} et~al.}{2013}]{Foreman-Mackey2013}
{Foreman-Mackey} D.,  {Hogg} D.~W.,  {Lang} D.,   {Goodman} J.,  2013, \mn@doi
  [\pasp] {10.1086/670067}, \href
  {https://ui.adsabs.harvard.edu/abs/2013PASP..125..306F} {125, 306}

\bibitem[\protect\citeauthoryear{{Freedman}}{{Freedman}}{2021}]{Freedman2021}
{Freedman} W.~L.,  2021, \mn@doi [\apj] {10.3847/1538-4357/ac0e95}, \href
  {https://ui.adsabs.harvard.edu/abs/2021ApJ...919...16F} {919, 16}

\bibitem[\protect\citeauthoryear{{Freedman} et~al.,}{{Freedman}
  et~al.}{2019}]{Freedman2019}
{Freedman} W.~L.,  et~al., 2019, \mn@doi [\apj] {10.3847/1538-4357/ab2f73},
  \href {https://ui.adsabs.harvard.edu/abs/2019ApJ...882...34F} {882, 34}

\bibitem[\protect\citeauthoryear{{Friedman} et~al.,}{{Friedman}
  et~al.}{2015}]{Friedman2015}
{Friedman} A.~S.,  et~al., 2015, \mn@doi [\apjs] {10.1088/0067-0049/220/1/9},
  \href {https://ui.adsabs.harvard.edu/abs/2015ApJS..220....9F} {220, 9}

\bibitem[\protect\citeauthoryear{{Galbany} et~al.,}{{Galbany}
  et~al.}{2022}]{galbany2022}
{Galbany} L.,  et~al., 2022, arXiv e-prints, \href
  {https://ui.adsabs.harvard.edu/abs/2022arXiv220902546G} {p. arXiv:2209.02546}

\bibitem[\protect\citeauthoryear{{Gall} et~al.,}{{Gall}
  et~al.}{2018}]{Gall2018}
{Gall} C.,  et~al., 2018, \mn@doi [\aap] {10.1051/0004-6361/201730886}, \href
  {https://ui.adsabs.harvard.edu/abs/2018A&A...611A..58G} {611, A58}

\bibitem[\protect\citeauthoryear{{Guy} et~al.,}{{Guy} et~al.}{2007}]{Guy2007}
{Guy} J.,  et~al., 2007, \mn@doi [\aap] {10.1051/0004-6361:20066930}, \href
  {http://adsabs.harvard.edu/abs/2007A%26A...466...11G} {466, 11}

\bibitem[\protect\citeauthoryear{{Hamuy}, {Phillips}, {Maza}, {Wischnjewsky},
  {Uomoto}, {Landolt}  \& {Khatwani}}{{Hamuy} et~al.}{1991}]{Hamuy1991}
{Hamuy} M.,  {Phillips} M.~M.,  {Maza} J.,  {Wischnjewsky} M.,  {Uomoto} A.,
  {Landolt} A.~U.,   {Khatwani} R.,  1991, \mn@doi [\aj] {10.1086/115867},
  \href {https://ui.adsabs.harvard.edu/abs/1991AJ....102..208H} {102, 208}

\bibitem[\protect\citeauthoryear{Hoffman \& Gelman}{Hoffman \&
  Gelman}{2014}]{Hoffman14}
Hoffman M.~D.,  Gelman A.,  2014, J. Machine Learning Res., 15, 1593

\bibitem[\protect\citeauthoryear{{Hounsell} et~al.,}{{Hounsell}
  et~al.}{2018}]{hounsell2018}
{Hounsell} R.,  et~al., 2018, \mn@doi [\apj] {10.3847/1538-4357/aac08b}, \href
  {https://ui.adsabs.harvard.edu/abs/2018ApJ...867...23H} {867, 23}

\bibitem[\protect\citeauthoryear{{Jha} et~al.,}{{Jha} et~al.}{1999}]{Jha1999}
{Jha} S.,  et~al., 1999, \mn@doi [\apjs] {10.1086/313275}, \href
  {https://ui.adsabs.harvard.edu/abs/1999ApJS..125...73J} {125, 73}

\bibitem[\protect\citeauthoryear{{Jha} et~al.,}{{Jha} et~al.}{2019}]{Jha2019}
{Jha} S.~W.,  et~al., 2019, {Supernovae in the Infrared avec Hubble}, HST
  Proposal. Cycle 27, ID. \#15889

\bibitem[\protect\citeauthoryear{{Johansson} et~al.,}{{Johansson}
  et~al.}{2021}]{Johansson2021}
{Johansson} J.,  et~al., 2021, \mn@doi [\apj] {10.3847/1538-4357/ac2f9e}, \href
  {https://ui.adsabs.harvard.edu/abs/2021ApJ...923..237J} {923, 237}

\bibitem[\protect\citeauthoryear{{Jones} et~al.,}{{Jones}
  et~al.}{2018}]{Jones2018}
{Jones} D.~O.,  et~al., 2018, \mn@doi [\apj] {10.3847/1538-4357/aae2b9}, \href
  {https://ui.adsabs.harvard.edu/abs/2018ApJ...867..108J} {867, 108}

\bibitem[\protect\citeauthoryear{{Jones} et~al.,}{{Jones}
  et~al.}{2021}]{Jones2021}
{Jones} D.~O.,  et~al., 2021, \mn@doi [\apj] {10.3847/1538-4357/abd7f5}, \href
  {https://ui.adsabs.harvard.edu/abs/2021ApJ...908..143J} {908, 143}

\bibitem[\protect\citeauthoryear{{Jones} et~al.,}{{Jones}
  et~al.}{2022}]{Jones2022:raisin}
{Jones} D.~O.,  et~al., 2022, \mn@doi [\apj] {10.3847/1538-4357/ac755b}, \href
  {https://ui.adsabs.harvard.edu/abs/2022ApJ...933..172J} {933, 172}

\bibitem[\protect\citeauthoryear{{Kattner} et~al.,}{{Kattner}
  et~al.}{2012}]{Kattner2012}
{Kattner} S.,  et~al., 2012, \mn@doi [\pasp] {10.1086/664734}, \href
  {https://ui.adsabs.harvard.edu/abs/2012PASP..124..114K} {124, 114}

\bibitem[\protect\citeauthoryear{{Kenworthy} et~al.,}{{Kenworthy}
  et~al.}{2022}]{Kenworthy2022}
{Kenworthy} W.~D.,  et~al., 2022, \mn@doi [\apj] {10.3847/1538-4357/ac80bd},
  \href {https://ui.adsabs.harvard.edu/abs/2022ApJ...935...83K} {935, 83}

\bibitem[\protect\citeauthoryear{{Knox} \& {Millea}}{{Knox} \&
  {Millea}}{2020}]{2020PhRvD.101d3533K}
{Knox} L.,  {Millea} M.,  2020, \mn@doi [\prd] {10.1103/PhysRevD.101.043533},
  \href {https://ui.adsabs.harvard.edu/abs/2020PhRvD.101d3533K} {101, 043533}

\bibitem[\protect\citeauthoryear{{Konchady}, {Oelkers}, {Jones}, {Yuan},
  {Macri}, {Peterson}  \& {Riess}}{{Konchady} et~al.}{2022}]{Konchady2022}
{Konchady} T.,  {Oelkers} R.~J.,  {Jones} D.~O.,  {Yuan} W.,  {Macri} L.~M.,
  {Peterson} E.~R.,   {Riess} A.~G.,  2022, \mn@doi [\apjs]
  {10.3847/1538-4365/ac41d3}, \href
  {https://ui.adsabs.harvard.edu/abs/2022ApJS..258...24K} {258, 24}

\bibitem[\protect\citeauthoryear{{Krisciunas}, {Hastings}, {Loomis},
  {McMillan}, {Rest}, {Riess}  \& {Stubbs}}{{Krisciunas}
  et~al.}{2000}]{Krisciunas2000}
{Krisciunas} K.,  {Hastings} N.~C.,  {Loomis} K.,  {McMillan} R.,  {Rest} A.,
  {Riess} A.~G.,   {Stubbs} C.,  2000, \mn@doi [\apj] {10.1086/309263}, \href
  {https://ui.adsabs.harvard.edu/abs/2000ApJ...539..658K} {539, 658}

\bibitem[\protect\citeauthoryear{{Krisciunas} et~al.,}{{Krisciunas}
  et~al.}{2003}]{Krisciunas2003}
{Krisciunas} K.,  et~al., 2003, \mn@doi [\aj] {10.1086/345571}, \href
  {https://ui.adsabs.harvard.edu/abs/2003AJ....125..166K} {125, 166}

\bibitem[\protect\citeauthoryear{{Krisciunas}, {Phillips}  \&
  {Suntzeff}}{{Krisciunas} et~al.}{2004}]{krisciunas2004}
{Krisciunas} K.,  {Phillips} M.~M.,   {Suntzeff} N.~B.,  2004, \mn@doi [\apjl]
  {10.1086/382731}, \href
  {https://ui.adsabs.harvard.edu/abs/2004ApJ...602L..81K} {602, L81}

\bibitem[\protect\citeauthoryear{{Krisciunas} et~al.,}{{Krisciunas}
  et~al.}{2007}]{Krisciunas2007}
{Krisciunas} K.,  et~al., 2007, \mn@doi [\aj] {10.1086/509126}, \href
  {https://ui.adsabs.harvard.edu/abs/2007AJ....133...58K} {133, 58}

\bibitem[\protect\citeauthoryear{{Krisciunas} et~al.,}{{Krisciunas}
  et~al.}{2017}]{Krisciunas2017}
{Krisciunas} K.,  et~al., 2017, \mn@doi [\aj] {10.3847/1538-3881/aa8df0}, \href
  {https://ui.adsabs.harvard.edu/abs/2017AJ....154..211K} {154, 211}

\bibitem[\protect\citeauthoryear{{Leibundgut}, {Kirshner}, {Filippenko},
  {Shields}, {Foltz}, {Phillips}  \& {Sonneborn}}{{Leibundgut}
  et~al.}{1991}]{Leibundgut1991}
{Leibundgut} B.,  {Kirshner} R.~P.,  {Filippenko} A.~V.,  {Shields} J.~C.,
  {Foltz} C.~B.,  {Phillips} M.~M.,   {Sonneborn} G.,  1991, \mn@doi [\apjl]
  {10.1086/185993}, \href
  {https://ui.adsabs.harvard.edu/abs/1991ApJ...371L..23L} {371, L23}

\bibitem[\protect\citeauthoryear{{Li}, {Casertano}  \& {Riess}}{{Li}
  et~al.}{2022}]{Li2022}
{Li} S.,  {Casertano} S.,   {Riess} A.~G.,  2022, \mn@doi [\apj]
  {10.3847/1538-4357/ac7559}, \href
  {https://ui.adsabs.harvard.edu/abs/2022ApJ...939...96L} {939, 96}

\bibitem[\protect\citeauthoryear{{Lu} et~al.,}{{Lu} et~al.}{2022}]{Lu2022}
{Lu} J.,  et~al., 2022, arXiv e-prints, \href
  {https://ui.adsabs.harvard.edu/abs/2022arXiv221105998L} {p. arXiv:2211.05998}

\bibitem[\protect\citeauthoryear{{Mandel}, {Wood-Vasey}, {Friedman}  \&
  {Kirshner}}{{Mandel} et~al.}{2009}]{Mandel2009}
{Mandel} K.~S.,  {Wood-Vasey} W.~M.,  {Friedman} A.~S.,   {Kirshner} R.~P.,
  2009, \mn@doi [\apj] {10.1088/0004-637X/704/1/629}, \href
  {https://ui.adsabs.harvard.edu/abs/2009ApJ...704..629M} {704, 629}

\bibitem[\protect\citeauthoryear{{Mandel}, {Narayan}  \& {Kirshner}}{{Mandel}
  et~al.}{2011}]{Mandel2011}
{Mandel} K.~S.,  {Narayan} G.,   {Kirshner} R.~P.,  2011, \mn@doi [\apj]
  {10.1088/0004-637X/731/2/120}, \href
  {https://ui.adsabs.harvard.edu/abs/2011ApJ...731..120M} {731, 120}

\bibitem[\protect\citeauthoryear{{Mandel}, {Thorp}, {Narayan}, {Friedman}  \&
  {Avelino}}{{Mandel} et~al.}{2022}]{Mandel2020}
{Mandel} K.~S.,  {Thorp} S.,  {Narayan} G.,  {Friedman} A.~S.,   {Avelino} A.,
  2022, \mn@doi [\mnras] {10.1093/mnras/stab3496}, \href
  {https://ui.adsabs.harvard.edu/abs/2022MNRAS.510.3939M} {510, 3939}

\bibitem[\protect\citeauthoryear{{Marion} et~al.,}{{Marion}
  et~al.}{2016}]{Marion2016}
{Marion} G.~H.,  et~al., 2016, \mn@doi [\apj] {10.3847/0004-637X/820/2/92},
  \href {https://ui.adsabs.harvard.edu/abs/2016ApJ...820...92M} {820, 92}

\bibitem[\protect\citeauthoryear{{Matheson} et~al.,}{{Matheson}
  et~al.}{2012}]{Matheson2012}
{Matheson} T.,  et~al., 2012, \mn@doi [\apj] {10.1088/0004-637X/754/1/19},
  \href {https://ui.adsabs.harvard.edu/abs/2012ApJ...754...19M} {754, 19}

\bibitem[\protect\citeauthoryear{{Meikle}}{{Meikle}}{2000}]{meikle2000}
{Meikle} W.~P.~S.,  2000, \mn@doi [\mnras] {10.1046/j.1365-8711.2000.03411.x},
  \href {https://ui.adsabs.harvard.edu/abs/2000MNRAS.314..782M} {314, 782}

\bibitem[\protect\citeauthoryear{{M{\"o}rtsell}, {Goobar}, {Johansson}  \&
  {Dhawan}}{{M{\"o}rtsell} et~al.}{2022a}]{Mortsell2021a}
{M{\"o}rtsell} E.,  {Goobar} A.,  {Johansson} J.,   {Dhawan} S.,  2022a,
  \mn@doi [\apj] {10.3847/1538-4357/ac756e}, \href
  {https://ui.adsabs.harvard.edu/abs/2022ApJ...933..212M} {933, 212}

\bibitem[\protect\citeauthoryear{{M{\"o}rtsell}, {Goobar}, {Johansson}  \&
  {Dhawan}}{{M{\"o}rtsell} et~al.}{2022b}]{Mortsell2021b}
{M{\"o}rtsell} E.,  {Goobar} A.,  {Johansson} J.,   {Dhawan} S.,  2022b,
  \mn@doi [\apj] {10.3847/1538-4357/ac7c19}, \href
  {https://ui.adsabs.harvard.edu/abs/2022ApJ...935...58M} {935, 58}

\bibitem[\protect\citeauthoryear{{M{\"u}ller-Bravo} et~al.,}{{M{\"u}ller-Bravo}
  et~al.}{2022}]{Muller-Bravo2022}
{M{\"u}ller-Bravo} T.~E.,  et~al., 2022, \mn@doi [\aap]
  {10.1051/0004-6361/202243845}, \href
  {https://ui.adsabs.harvard.edu/abs/2022A&A...665A.123M} {665, A123}

\bibitem[\protect\citeauthoryear{{Peterson} et~al.,}{{Peterson}
  et~al.}{2021}]{Peterson2021}
{Peterson} E.~R.,  et~al., 2021, arXiv e-prints, \href
  {https://ui.adsabs.harvard.edu/abs/2021arXiv211003487P} {p. arXiv:2110.03487}

\bibitem[\protect\citeauthoryear{{Phillips}}{{Phillips}}{2012}]{phillips2012}
{Phillips} M.~M.,  2012, \mn@doi [\pasa] {10.1071/AS11056}, \href
  {https://ui.adsabs.harvard.edu/abs/2012PASA...29..434P} {29, 434}

\bibitem[\protect\citeauthoryear{{Phillips} et~al.,}{{Phillips}
  et~al.}{2019}]{Phillips2019}
{Phillips} M.~M.,  et~al., 2019, \mn@doi [\pasp] {10.1088/1538-3873/aae8bd},
  \href {https://ui.adsabs.harvard.edu/abs/2019PASP..131a4001P} {131, 014001}

\bibitem[\protect\citeauthoryear{{Planck Collaboration}}{{Planck
  Collaboration}}{2020}]{2020A&A...641A...6P}
{Planck Collaboration} 2020, \mn@doi [\aap] {10.1051/0004-6361/201833910},
  \href {https://ui.adsabs.harvard.edu/abs/2020A&A...641A...6P} {641, A6}

\bibitem[\protect\citeauthoryear{{Richmond} \& {Smith}}{{Richmond} \&
  {Smith}}{2012}]{Richmond2012}
{Richmond} M.~W.,  {Smith} H.~A.,  2012, JAAVSO, \href
  {https://ui.adsabs.harvard.edu/abs/2012JAVSO..40..872R} {40, 872}

\bibitem[\protect\citeauthoryear{{Richmond} et~al.,}{{Richmond}
  et~al.}{1995}]{Richmond1995}
{Richmond} M.~W.,  et~al., 1995, \mn@doi [\aj] {10.1086/117437}, \href
  {https://ui.adsabs.harvard.edu/abs/1995AJ....109.2121R} {109, 2121}

\bibitem[\protect\citeauthoryear{{Riess} et~al.,}{{Riess}
  et~al.}{1999}]{riess1999}
{Riess} A.~G.,  et~al., 1999, \mn@doi [\aj] {10.1086/301143}, \href
  {https://ui.adsabs.harvard.edu/abs/1999AJ....118.2675R} {118, 2675}

\bibitem[\protect\citeauthoryear{{Riess} et~al.,}{{Riess}
  et~al.}{2005}]{riess2005}
{Riess} A.~G.,  et~al., 2005, \mn@doi [\apj] {10.1086/430497}, \href
  {https://ui.adsabs.harvard.edu/abs/2005ApJ...627..579R} {627, 579}

\bibitem[\protect\citeauthoryear{{Riess} et~al.,}{{Riess}
  et~al.}{2016}]{riess2016}
{Riess} A.~G.,  et~al., 2016, \mn@doi [\apj] {10.3847/0004-637X/826/1/56},
  \href {https://ui.adsabs.harvard.edu/abs/2016ApJ...826...56R} {826, 56}

\bibitem[\protect\citeauthoryear{{Riess} et~al.,}{{Riess}
  et~al.}{2022}]{riess2021}
{Riess} A.~G.,  et~al., 2022, \mn@doi [\apjl] {10.3847/2041-8213/ac5c5b}, \href
  {https://ui.adsabs.harvard.edu/abs/2022ApJ...934L...7R} {934, L7}

\bibitem[\protect\citeauthoryear{{Rigault} et~al.,}{{Rigault}
  et~al.}{2020}]{Rigault2020}
{Rigault} M.,  et~al., 2020, \mn@doi [\aap] {10.1051/0004-6361/201730404},
  \href {https://ui.adsabs.harvard.edu/abs/2020A&A...644A.176R} {644, A176}

\bibitem[\protect\citeauthoryear{{Rose} et~al.,}{{Rose}
  et~al.}{2021}]{Rose2021}
{Rose} B.~M.,  et~al., 2021, arXiv e-prints, \href
  {https://ui.adsabs.harvard.edu/abs/2021arXiv211103081R} {p. arXiv:2111.03081}

\bibitem[\protect\citeauthoryear{{Sch{\"o}neberg}, {Abell{\'a}n},
  {S{\'a}nchez}, {Witte}, {Poulin}  \& {Lesgourgues}}{{Sch{\"o}neberg}
  et~al.}{2022}]{Schonberg2022}
{Sch{\"o}neberg} N.,  {Abell{\'a}n} G.~F.,  {S{\'a}nchez} A.~P.,  {Witte}
  S.~J.,  {Poulin} V.,   {Lesgourgues} J.,  2022, \mn@doi [\physrep]
  {10.1016/j.physrep.2022.07.001}, \href
  {https://ui.adsabs.harvard.edu/abs/2022PhR...984....1S} {984, 1}

\bibitem[\protect\citeauthoryear{{Schweizer} et~al.,}{{Schweizer}
  et~al.}{2008}]{Schweizer2008}
{Schweizer} F.,  et~al., 2008, \mn@doi [\aj] {10.1088/0004-6256/136/4/1482},
  \href {https://ui.adsabs.harvard.edu/abs/2008AJ....136.1482S} {136, 1482}

\bibitem[\protect\citeauthoryear{{Scolnic} et~al.,}{{Scolnic}
  et~al.}{2021}]{scolnic2021}
{Scolnic} D.,  et~al., 2021, arXiv e-prints, \href
  {https://ui.adsabs.harvard.edu/abs/2021arXiv211203863S} {p. arXiv:2112.03863}

\bibitem[\protect\citeauthoryear{{Shah}, {Lemos}  \& {Lahav}}{{Shah}
  et~al.}{2021}]{Shah2021}
{Shah} P.,  {Lemos} P.,   {Lahav} O.,  2021, \mn@doi [\aapr]
  {10.1007/s00159-021-00137-4}, \href
  {https://ui.adsabs.harvard.edu/abs/2021A&ARv..29....9S} {29, 9}

\bibitem[\protect\citeauthoryear{{Silverman} et~al.,}{{Silverman}
  et~al.}{2012}]{Silverman2012}
{Silverman} J.~M.,  et~al., 2012, \mn@doi [\mnras]
  {10.1111/j.1365-2966.2012.21270.x}, \href
  {https://ui.adsabs.harvard.edu/abs/2012MNRAS.425.1789S} {425, 1789}

\bibitem[\protect\citeauthoryear{{Stan Development Team}}{{Stan Development
  Team}}{2020}]{Stan2020}
{Stan Development Team} 2020, Stan Modelling Language Users Guide and Reference
  Manual v.2.25.
\url {https://mc-stan.org}

\bibitem[\protect\citeauthoryear{{Stritzinger} et~al.,}{{Stritzinger}
  et~al.}{2010}]{Stritzinger2010}
{Stritzinger} M.,  et~al., 2010, \mn@doi [\aj] {10.1088/0004-6256/140/6/2036},
  \href {https://ui.adsabs.harvard.edu/abs/2010AJ....140.2036S} {140, 2036}

\bibitem[\protect\citeauthoryear{{Stritzinger} et~al.,}{{Stritzinger}
  et~al.}{2011}]{Stritzinger2011}
{Stritzinger} M.~D.,  et~al., 2011, \mn@doi [\aj]
  {10.1088/0004-6256/142/5/156}, \href
  {https://ui.adsabs.harvard.edu/abs/2011AJ....142..156S} {142, 156}

\bibitem[\protect\citeauthoryear{{Thorp} \& {Mandel}}{{Thorp} \&
  {Mandel}}{2022}]{Thorp2022}
{Thorp} S.,  {Mandel} K.~S.,  2022, \mn@doi [\mnras] {10.1093/mnras/stac2714},
  \href {https://ui.adsabs.harvard.edu/abs/2022MNRAS.517.2360T} {517, 2360}

\bibitem[\protect\citeauthoryear{{Thorp}, {Mandel}, {Jones}, {Ward}  \&
  {Narayan}}{{Thorp} et~al.}{2021}]{Thorp2021}
{Thorp} S.,  {Mandel} K.~S.,  {Jones} D.~O.,  {Ward} S.~M.,   {Narayan} G.,
  2021, \mn@doi [\mnras] {10.1093/mnras/stab2849}, \href
  {https://ui.adsabs.harvard.edu/abs/2021MNRAS.508.4310T} {508, 4310}

\bibitem[\protect\citeauthoryear{{Tsvetkov}}{{Tsvetkov}}{1982}]{Tsvetkov1982}
{Tsvetkov} D.~Y.,  1982, Soviet Astronomy Letters, \href
  {https://ui.adsabs.harvard.edu/abs/1982SvAL....8..115T} {8, 115}

\bibitem[\protect\citeauthoryear{{Uddin} et~al.,}{{Uddin}
  et~al.}{2020}]{uddin2020}
{Uddin} S.~A.,  et~al., 2020, \mn@doi [\apj] {10.3847/1538-4357/abafb7}, \href
  {https://ui.adsabs.harvard.edu/abs/2020ApJ...901..143U} {901, 143}

\bibitem[\protect\citeauthoryear{{Walker} \& {Marino}}{{Walker} \&
  {Marino}}{1982}]{Walker1982}
{Walker} W.~S.~G.,  {Marino} B.~F.,  1982, Royal Astronomical Society of New
  Zealand Publications of Variable Star Section, \href
  {https://ui.adsabs.harvard.edu/abs/1982PVSS...10...53W} {10, 53}

\bibitem[\protect\citeauthoryear{{Ward} et~al.,}{{Ward}
  et~al.}{2022}]{Ward2022}
{Ward} S.~M.,  et~al., 2022, arXiv e-prints, \href
  {https://ui.adsabs.harvard.edu/abs/2022arXiv220910558W} {p. arXiv:2209.10558}

\bibitem[\protect\citeauthoryear{{Wells} et~al.,}{{Wells}
  et~al.}{1994}]{Wells1994}
{Wells} L.~A.,  et~al., 1994, \mn@doi [\aj] {10.1086/117236}, \href
  {https://ui.adsabs.harvard.edu/abs/1994AJ....108.2233W} {108, 2233}

\bibitem[\protect\citeauthoryear{{Weyant}, {Wood-Vasey}, {Allen}, {Garnavich},
  {Jha}, {Joyce}  \& {Matheson}}{{Weyant} et~al.}{2014}]{Weyant2014}
{Weyant} A.,  {Wood-Vasey} W.~M.,  {Allen} L.,  {Garnavich} P.~M.,  {Jha}
  S.~W.,  {Joyce} R.,   {Matheson} T.,  2014, \mn@doi [\apj]
  {10.1088/0004-637X/784/2/105}, \href
  {https://ui.adsabs.harvard.edu/abs/2014ApJ...784..105W} {784, 105}

\bibitem[\protect\citeauthoryear{{Weyant} et~al.,}{{Weyant}
  et~al.}{2018}]{Weyant2018}
{Weyant} A.,  et~al., 2018, \mn@doi [\aj] {10.3847/1538-3881/aab901}, \href
  {https://ui.adsabs.harvard.edu/abs/2018AJ....155..201W} {155, 201}

\bibitem[\protect\citeauthoryear{{Wood-Vasey} et~al.,}{{Wood-Vasey}
  et~al.}{2008}]{Wood-Vasey2008}
{Wood-Vasey} W.~M.,  et~al., 2008, \mn@doi [\apj] {10.1086/592374}, \href
  {https://ui.adsabs.harvard.edu/abs/2008ApJ...689..377W} {689, 377}

\makeatother
\end{thebibliography}
